\documentclass{revtex4}
\usepackage{slashed}
\usepackage{bigints}
\usepackage{amsmath}
\usepackage{graphicx,subfigure}
\usepackage{titlesec}

\begin{document}
\title{Neutral pion photoproduction on protons in fully covariant ChPT with $\Delta(1232)$ loop contributions}

\author{Astrid Hiller Blin}
\affiliation{Departamento de F{\'i}sica 
  Te{\'o}rica, Universidad de Valencia and IFIC, 
  Centro Mixto Universidad de Valencia-CSIC, Institutos 
  de Investigaci{\'o}n de Paterna, Aptdo. 22085, 46071 Valencia, 
  Spain} 

\author{Tim Ledwig}
\affiliation{Departamento de F{\'i}sica 
  Te{\'o}rica, Universidad de Valencia and IFIC, 
  Centro Mixto Universidad de Valencia-CSIC, Institutos 
  de Investigaci{\'o}n de Paterna, Aptdo. 22085, 46071 Valencia, 
  Spain} 

\author{Manuel Vicente Vacas}
\affiliation{Departamento de F{\'i}sica 
  Te{\'o}rica, Universidad de Valencia and IFIC, 
  Centro Mixto Universidad de Valencia-CSIC, Institutos 
  de Investigaci{\'o}n de Paterna, Aptdo. 22085, 46071 Valencia, 
  Spain}

\begin{abstract}
We study the neutral pion photoproduction at near-threshold energies in fully covariant chiral perturbation theory up to $\mathcal{O}(p^3)$. 
When including only nucleonic virtual states in the model, the convergence is too slow. Therefore we test the model when introducing the $\Delta(1232)$ resonance as an additional degree of freedom. Some low-energy constants were fitted, converging 
to values in good agreement with those expected from literature.
\end{abstract}

\maketitle

\section{Introduction}

Pion photoproduction on nucleons has been the focus of many theoretical studies in the past years~\cite{DeBaenst:1971hp,Vainshtein:1972ih,
Drechsel:1992pn,Bernard:2006gx,Gasparyan:2010xz}. More specifically the neutral channels are particularly 
interesting, as their total cross sections are much smaller than for the charged channels, and there are big discrepancies 
between the models and data~\cite{Mazzucato:1986dz,Beck:1990da}. It was pointed out that, when using chiral perturbation theory (ChPT) models it is therefore very important to take 
into account not only tree-level, but also loop-diagram contributions, as the lower-order contributions show very strong cancellations between 
amplitude pieces~\cite{Bernard:1991rt,Bernard:1992nc,Bernard:1994gm,Bernard:1995cj,Bernard:2001gz}.

By obtaining data for higher energies~\cite{Hornidge:2012ca}, it became clear that also this approach is not sufficient to describe the right convergence, 
even at energies of about only 20~MeV above threshold. A possible solution that was performed in~\cite{FernandezRamirez:2012nw,Hilt:2013uf} 
is studying higher-order contributions, i.e. going to $\mathcal{O}(p^4)$. Unfortunately this was not sufficient to reproduce the empirical behaviour, neither in the heavy-baryon quasi-static 
approach nor in the fully relativistical one. 
Another suggestion to solve the discrepancy, which 
we study here, is to consider the contributions coming from the $\Delta(1232)$ resonance~\cite{Ericson:1988gk,Hemmert:1996xg,Pascalutsa:2006up}. Its effect is small close to pion production threshold, but when approaching 
the resonance's mass one would expect it to lead to an important modification of the cross-section's behaviour~\cite{Chew:1957tf,Adler:1968tw,Pascalutsa:2004pk,Pascalutsa:2005vq,FernandezRamirez:2005iv,Alarcon:2012kn}. A first such study has already been 
performed in~\cite{Blin:2015}, which is a calculation up to chiral order $p^3$. There it has been shown that, although no new fitting constants 
are introduced, the convergence between model and data is very good, even at energies higher than 200~MeV.

Here we study the low-energy constants. They are taken 
as fitting parameters and their convergence to the literature values is tested. Furthermore, we propose the inclusion of the next possible set of loop diagrams. 
It corresponds to an order $p^{7/2}$ calculation when following the counting of~\cite{Lensky:2009uv}, which is reasonable for photon energies 
sufficiently below the $\Delta(1232)$ resonance mass. The goal is to study the effect of this higher order on the modelling of polarization observables, which is expected to be small if the 
theory is convergent and the power counting consistent.

\section{Specifics of the chiral Lagrangian to calculate the $\gamma p \rightarrow p \pi^0$ channel}

We study the process in a fully covariant ChPT calculation, including 
nucleons, pions, photons and the $\Delta(1232)$ resonance. The power-counting scheme we 
use is the $\delta$ expansion introduced in~\cite{Lensky:2009uv}, where the small parameter 
$\delta=M_\Delta-m$ is treated as being of order $p^{1/2}$. So being, the diagrams are ordered 
by the rule
\begin{equation}
D=4L+\sum{kV_{k}}-2N_\pi-N_N-\frac{1}{2}N_\Delta,
\label{eqOrder}
\end{equation}
where $D$ is the order of the diagram, $L$ the number of loops, $V_k$ the number of vertices of 
a specific order $k$, and the $N_i$ are the three propagators of the respective hadrons. Following 
this counting scheme, for a calculation up to $\mathcal{O}(p^{7/2})$ we need the following pieces 
of the nucleonic Lagrangian:
\begin{align}
\nonumber\mathcal{L}_{\pi N}=\bar{\Psi}\Bigg\{
&\mathrm{i}\slashed{\mathrm{D}}-m+\frac{g_A}{2}\slashed{u}\gamma_5
+\frac{1}{8m}\left(
c_6 f_{\mu\nu}^+ + c_7\text{Tr}\left[f_{\mu\nu}^+\right]
\right)\sigma^{\mu\nu}
\\
\nonumber&+\left(\frac{\mathrm{i}}{2m}\varepsilon^{\mu\nu\alpha\beta}\left(d_8\text{Tr}\left[\tilde f_{\mu\nu}^+u_\alpha\right]
 +d_9\text{Tr}\left[f_{\mu\nu}^+
\right]u_\alpha
\right)+\text{h.c.}\right)
\mathrm{D}_\beta
\\
 &+\frac{\gamma^{\mu}\gamma_5}{2}\left(
d_{16}\text{Tr}\left[\chi_+\right]u_\mu
 +\mathrm{i}d_{18}\gamma^{\mu}\gamma_5[\mathrm{D}_\mu,\chi_-]
\right)
\Bigg\}\Psi+\dots,
\label{eqLag12}
\end{align}
with the definitions of~\cite{Fettes:2000gb}. While the low-energy constants $g_A$, $c_6$ and $c_7$ are already well determined 
at the order at which they appear, corrections to their values are expected when moving to higher orders and including the $\Delta(1232)$ resonance. 
As for the $d_8$ and $d_9$, they are mainly sensitive to the pion production processes, as they represent the third-order contact term. Fitting them to 
our model is therefore an important study of their values. Finally, the $d_{16}$ and $d_{18}$ have already been thoroughly studied in~\cite{Alarcon:2012kn} in the same ChPT approach, degrees of freedom and renormalization as in the present work. Therefore, it would be extremely interesting to 
compare the fitting results of these two works. 
The inclusion of the degrees of freedom of the isospin-3/2 $\Delta$ quadruplet requires the additional Lagrangian terms
\begin{align}
\mathcal{L}_{\Delta\pi N}=\mathrm{i}\bar{\Psi}\left(\frac{h_A}{2FM_\Delta}T^a\gamma^{\mu\nu\lambda}(\mathrm{D}_\lambda^{ab}\pi^b)
+\frac{3e}{2m(m+M_\Delta)}T^3\left(\mathrm{i}g_M\tilde{F}^ {\mu\nu}-g_E\gamma_5F^{\mu\nu}\right)\right)\partial_\mu\Delta_\nu + \text{H.c.},
\end{align}
with the definitions in~\cite{Pascalutsa:2002pi}. The constants $F$, $h_A$ and $g_M$ are very well studied and we don't expect them to change in 
our fits. As for $g_E$, it is still not too well known, so we can leave it as a free fitting parameter.

\begin{figure}
\begin{center}
\includegraphics[width=0.24\textwidth]{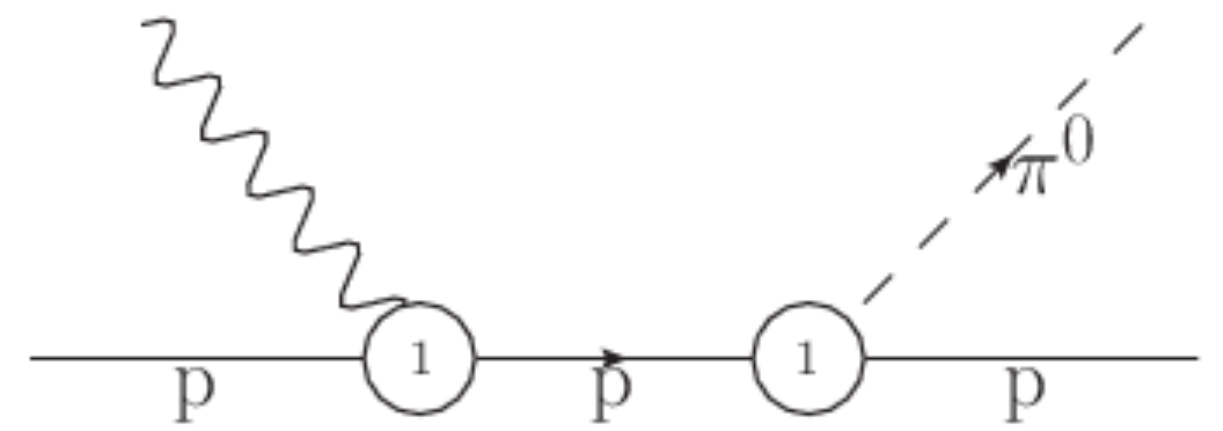}
\includegraphics[width=0.24\textwidth]{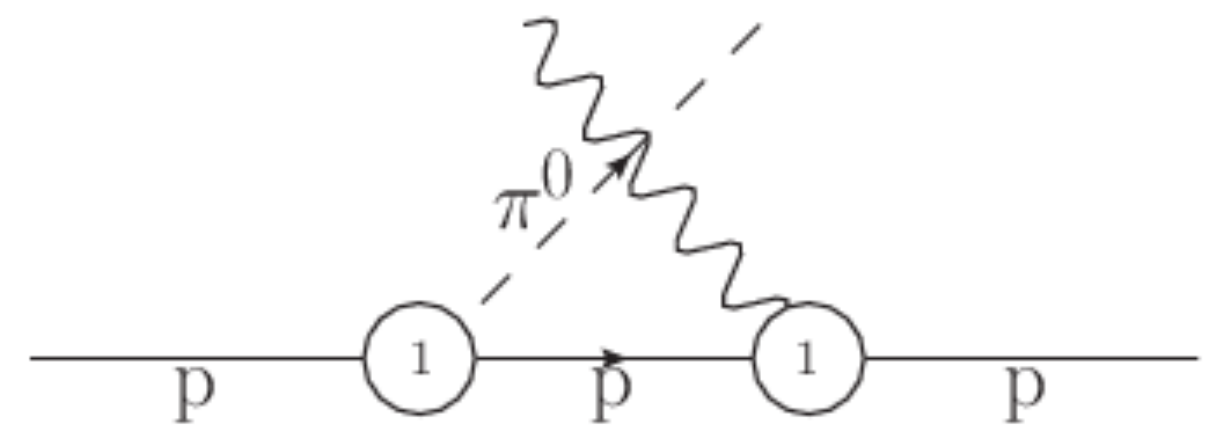}
\includegraphics[width=0.24\textwidth]{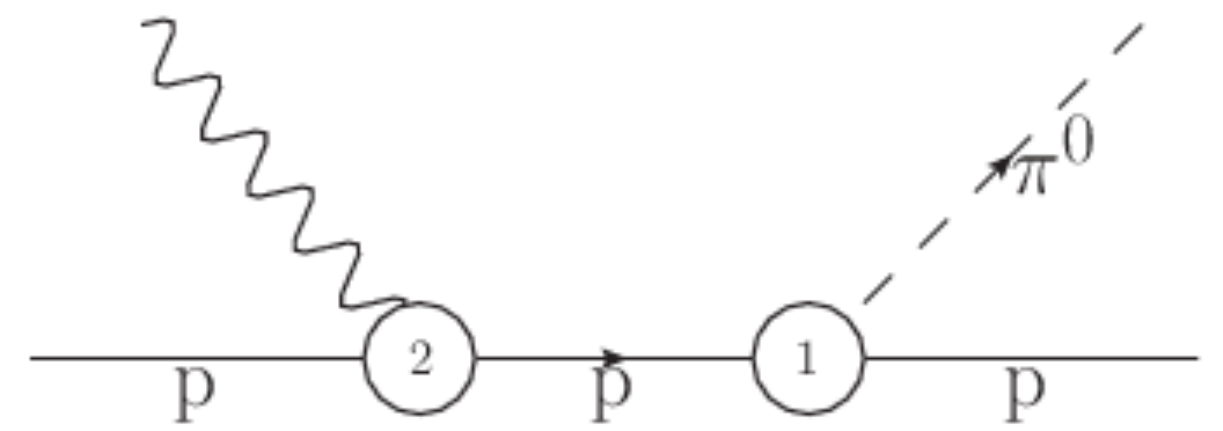}
\includegraphics[width=0.24\textwidth]{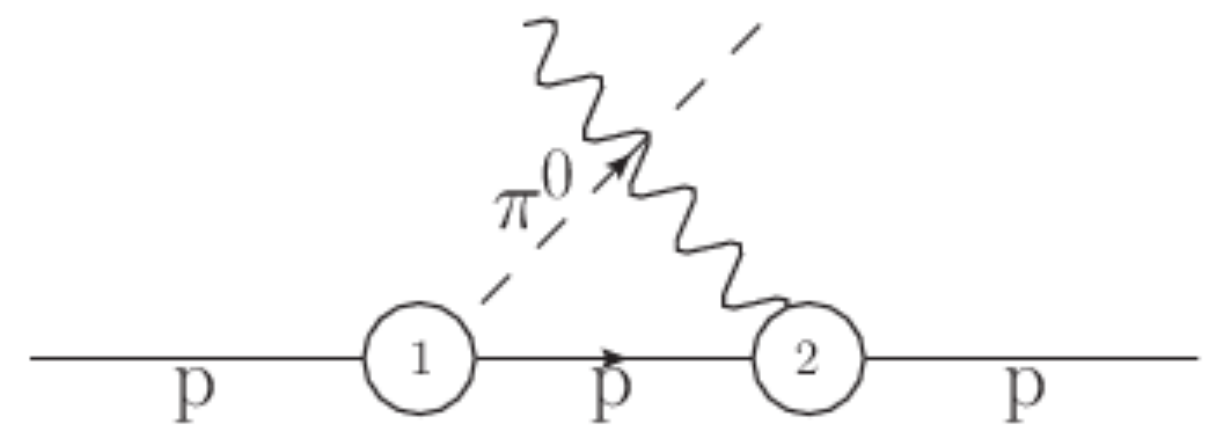}\vspace{5mm}\\
\includegraphics[width=0.3\textwidth]{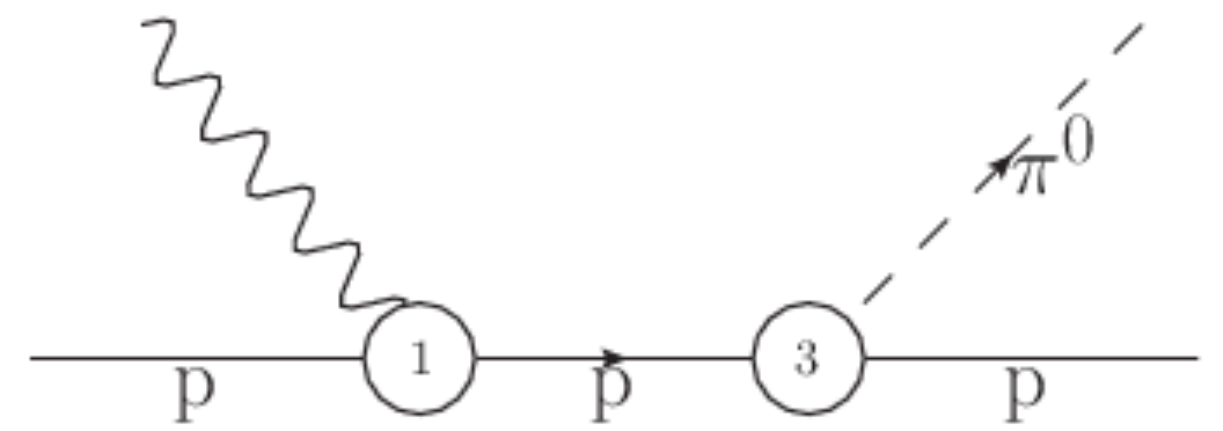}
\includegraphics[width=0.3\textwidth]{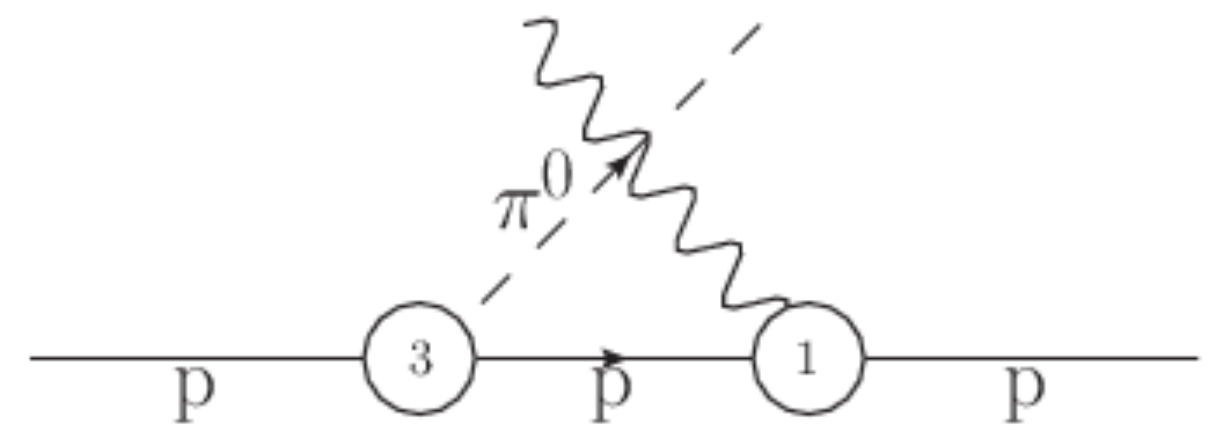}
\includegraphics[width=0.2\textwidth]{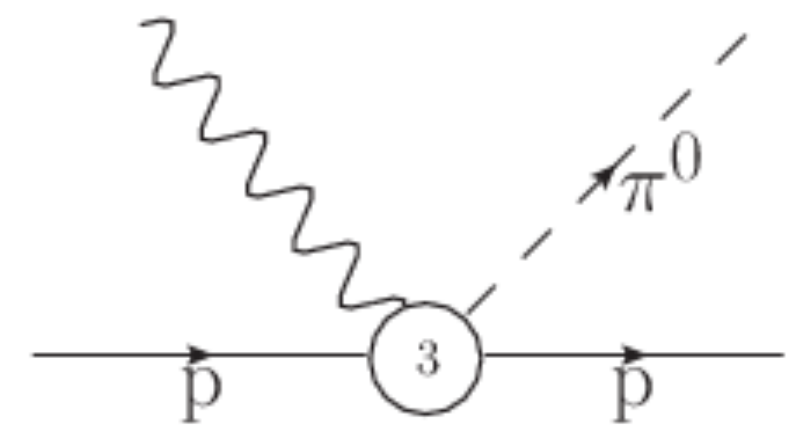}\vspace{5mm}\\
\includegraphics[width=0.2\textwidth]{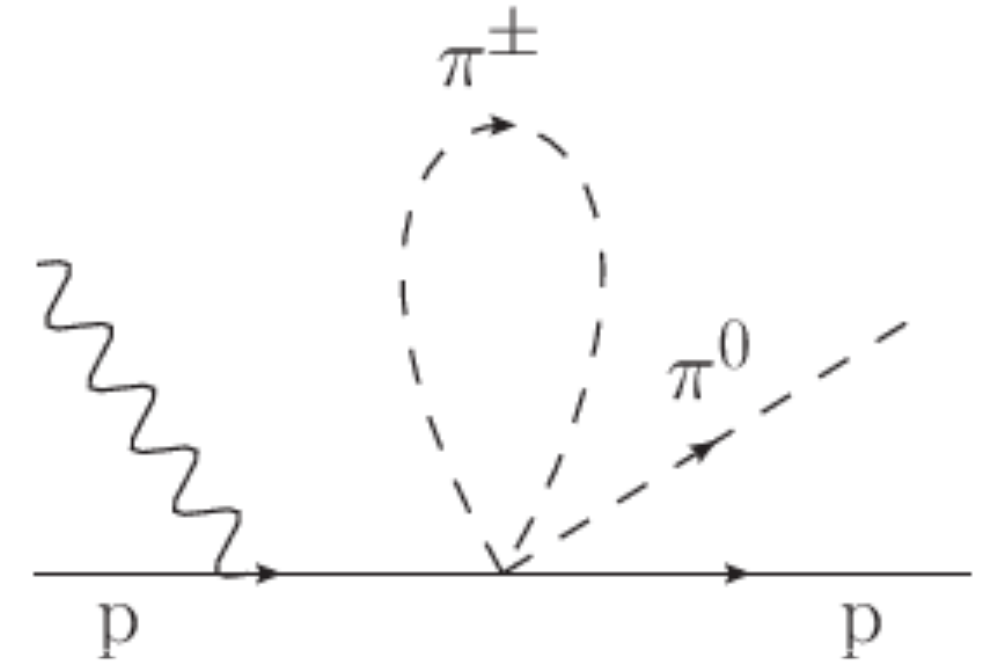}
\includegraphics[width=0.2\textwidth]{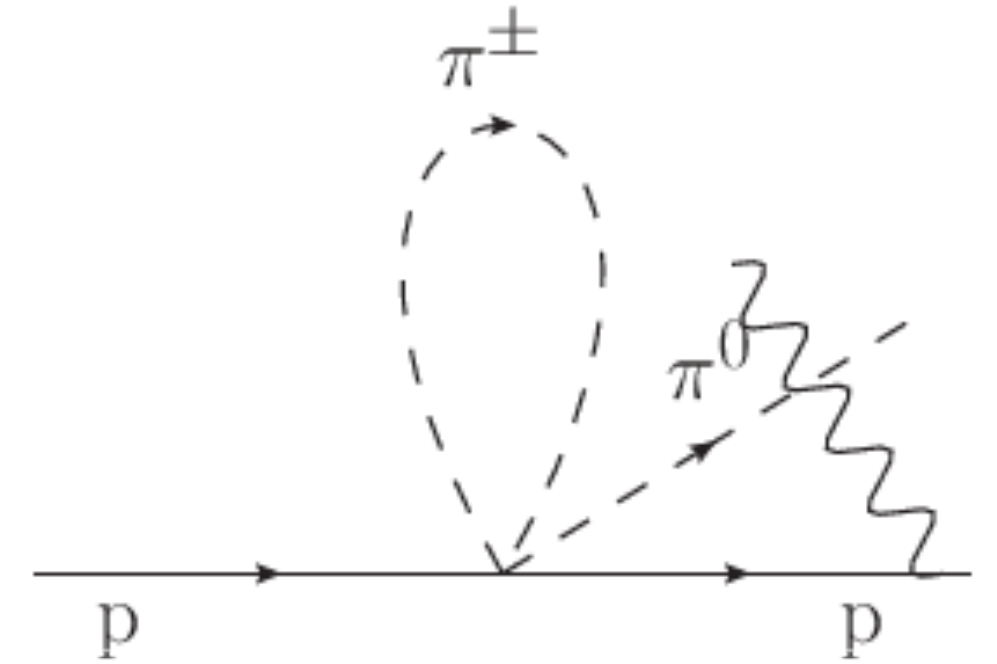}
\includegraphics[width=0.2\textwidth]{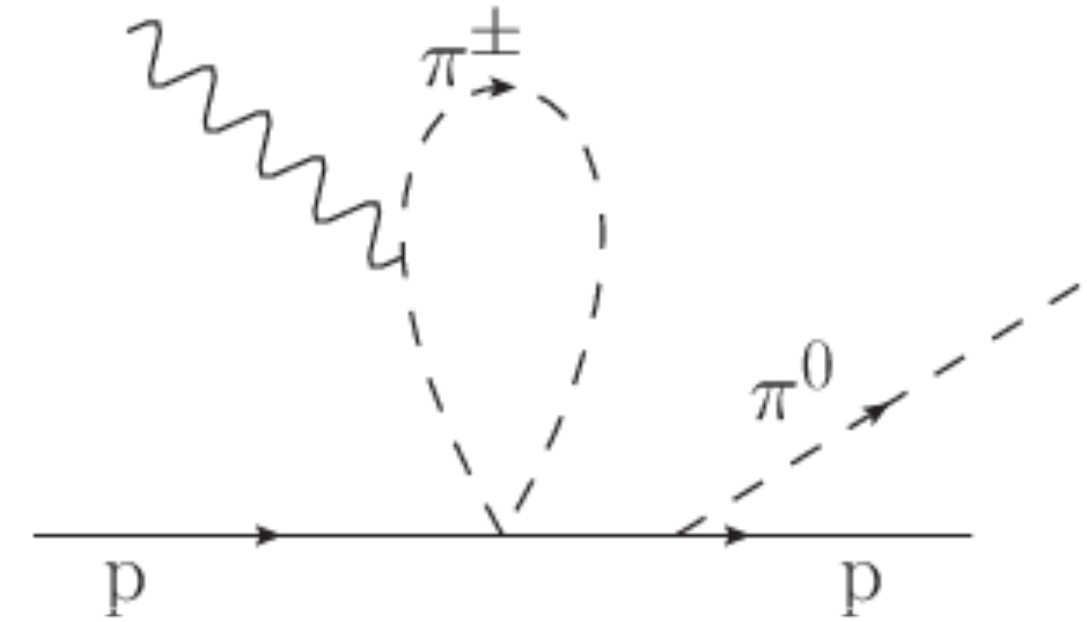}\vspace{5mm}\\
\includegraphics[width=0.2\textwidth]{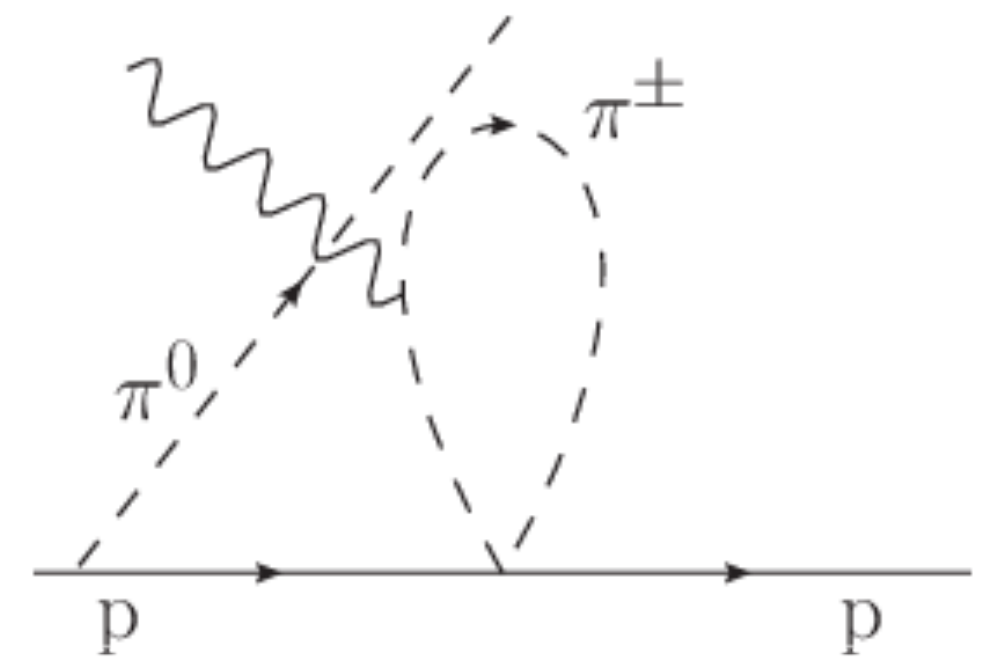}
\includegraphics[width=0.2\textwidth]{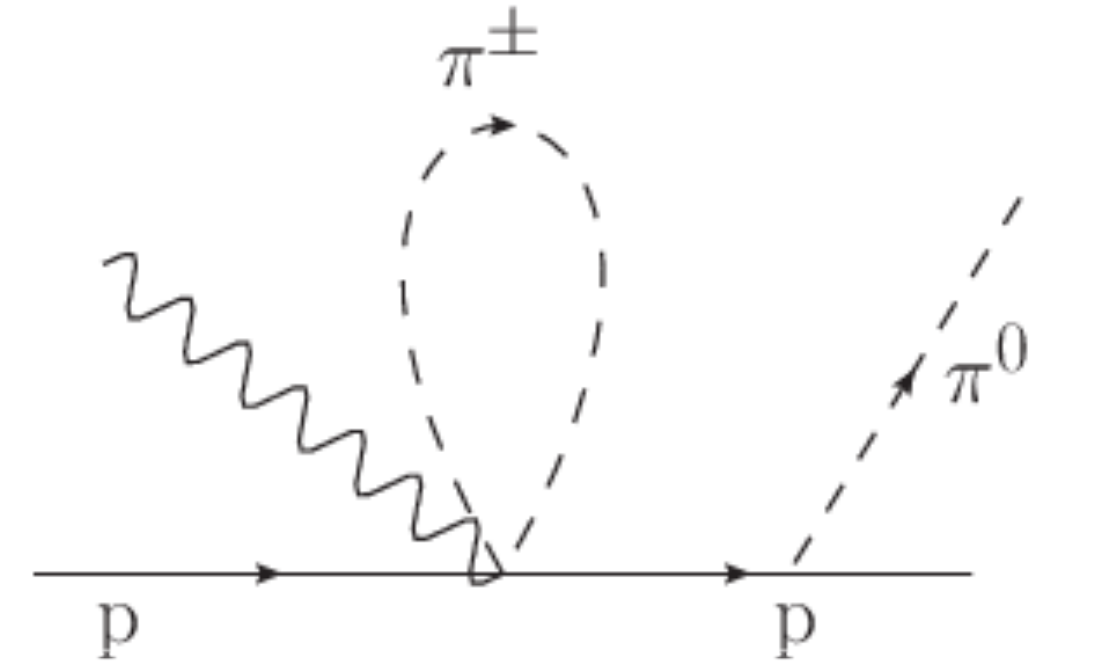}
\includegraphics[width=0.2\textwidth]{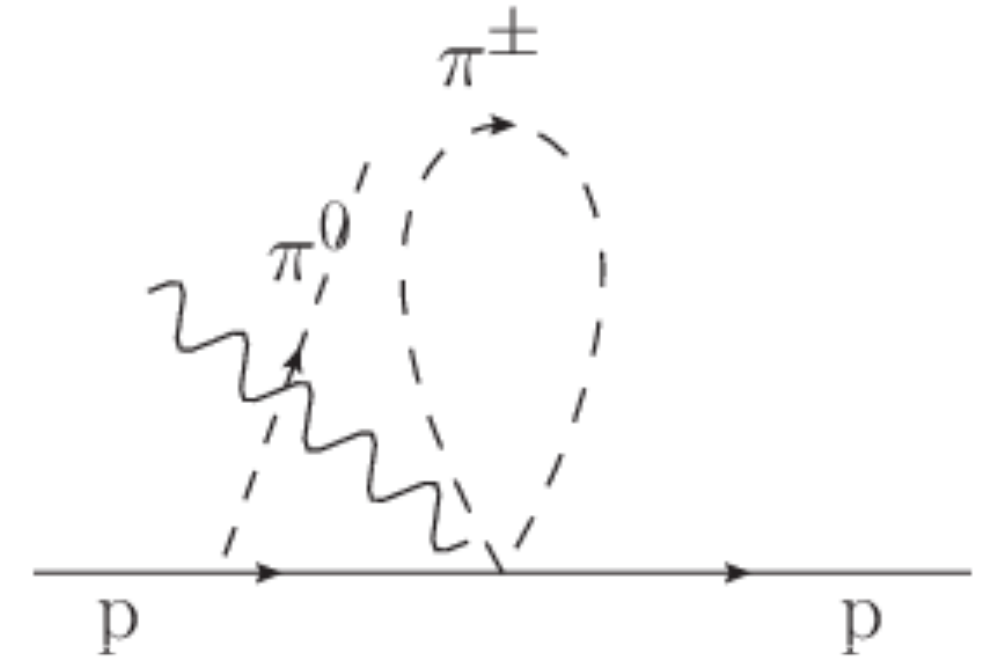}\vspace{5mm}\\
\includegraphics[width=0.2\textwidth]{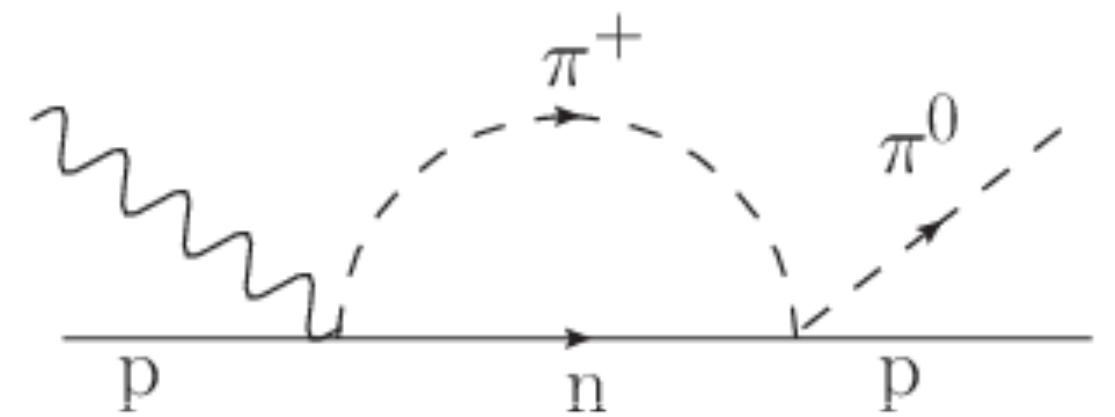}
\includegraphics[width=0.2\textwidth]{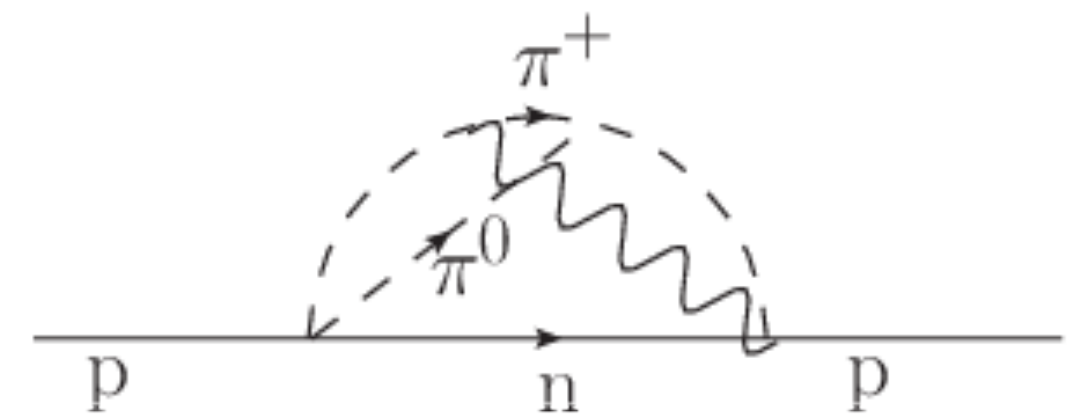}
\includegraphics[width=0.2\textwidth]{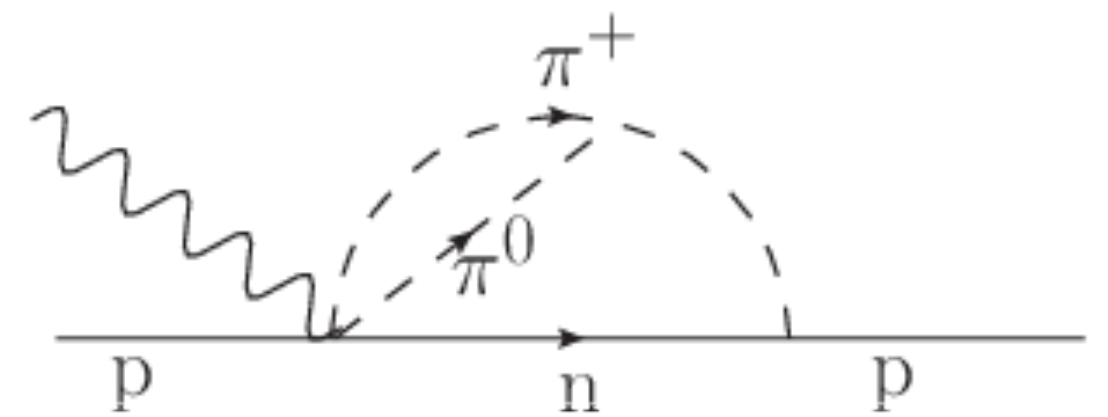}
\includegraphics[width=0.2\textwidth]{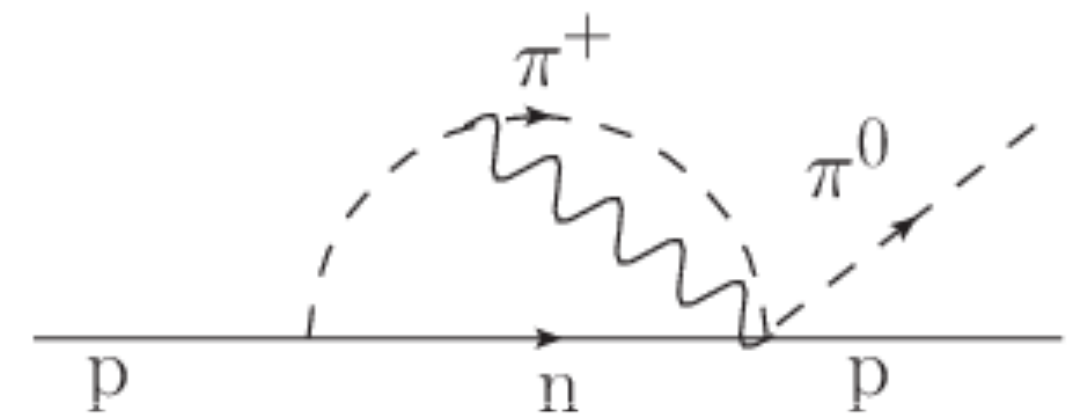}\vspace{5mm}\\
\includegraphics[width=0.2\textwidth]{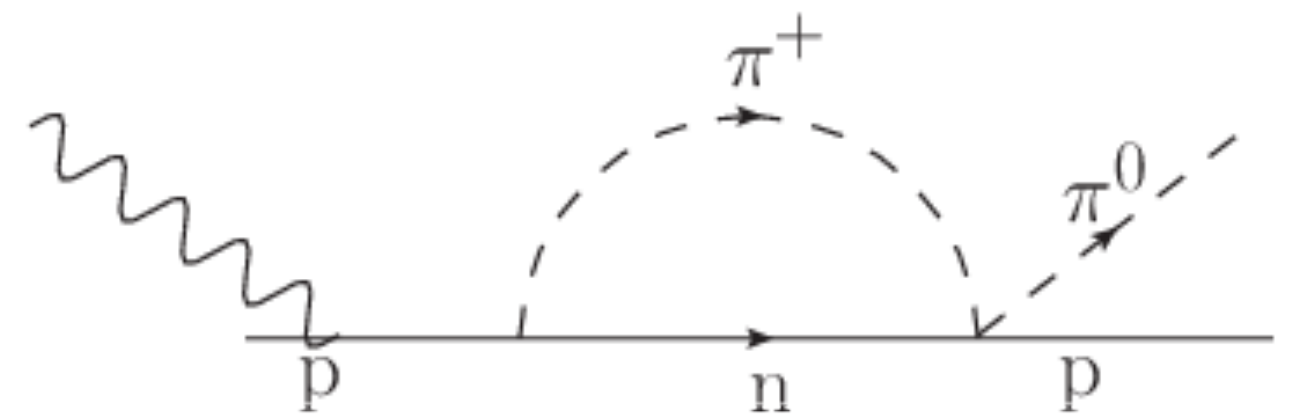}
\includegraphics[width=0.2\textwidth]{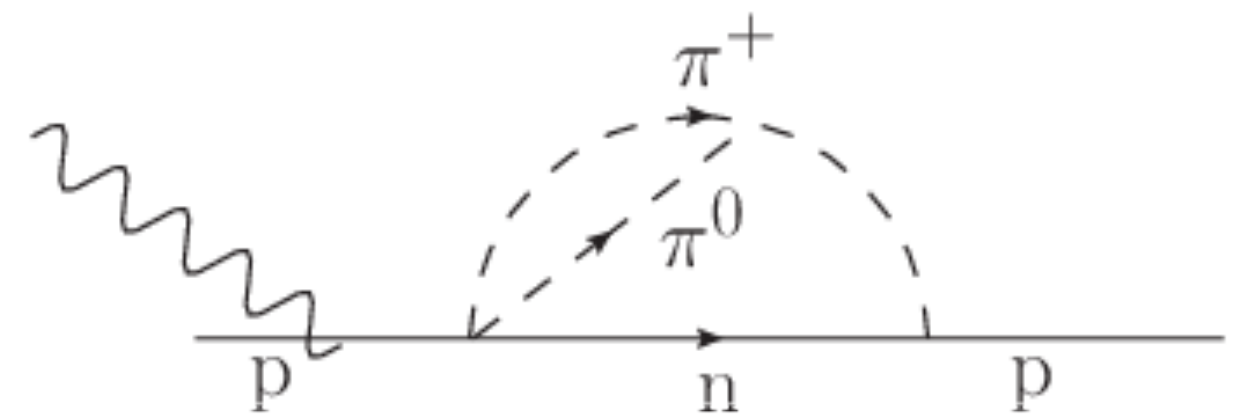}
\includegraphics[width=0.2\textwidth]{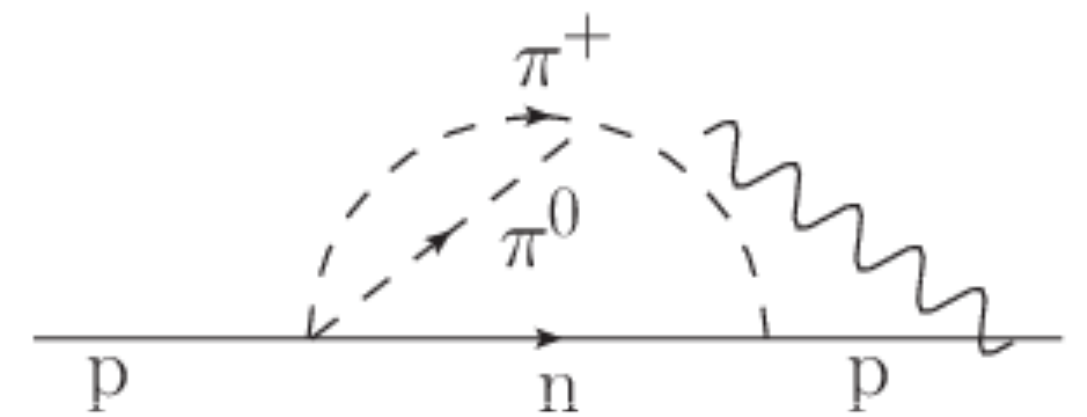}
\includegraphics[width=0.2\textwidth]{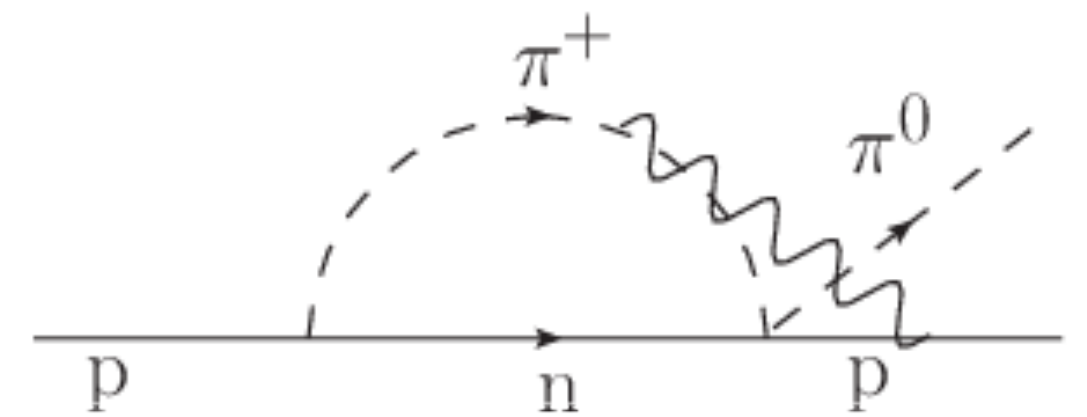}\\
\includegraphics[width=0.2\textwidth]{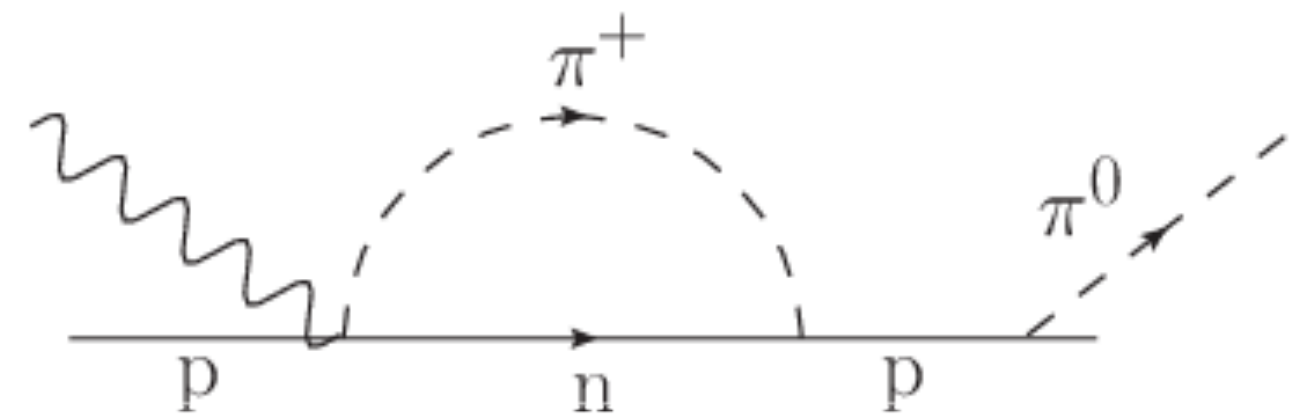}
\includegraphics[width=0.2\textwidth]{O3L2f.pdf}
\includegraphics[width=0.2\textwidth]{O3L2g.pdf}
\includegraphics[width=0.2\textwidth]{O3L2h.pdf}\vspace{5mm}\\
\includegraphics[width=0.2\textwidth]{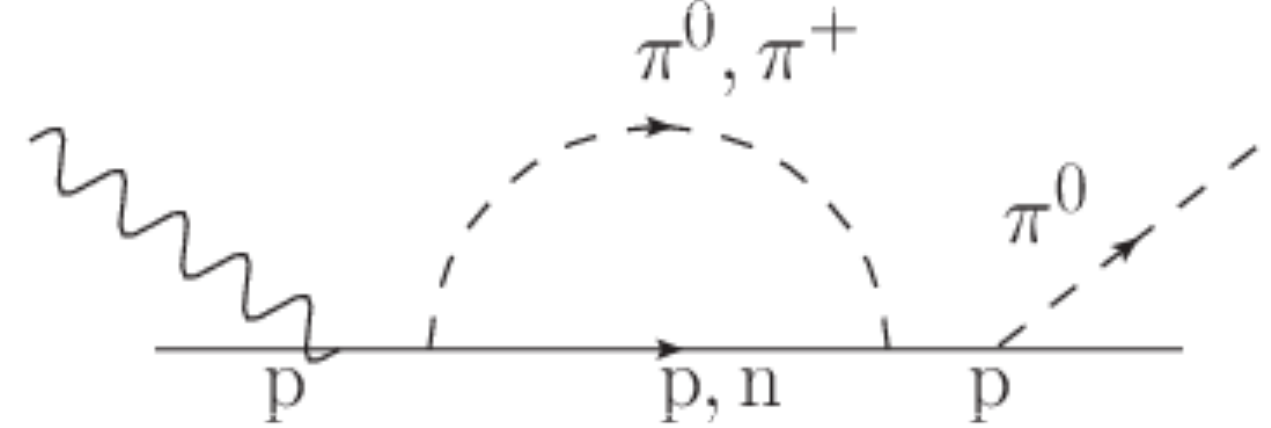}
\includegraphics[width=0.2\textwidth]{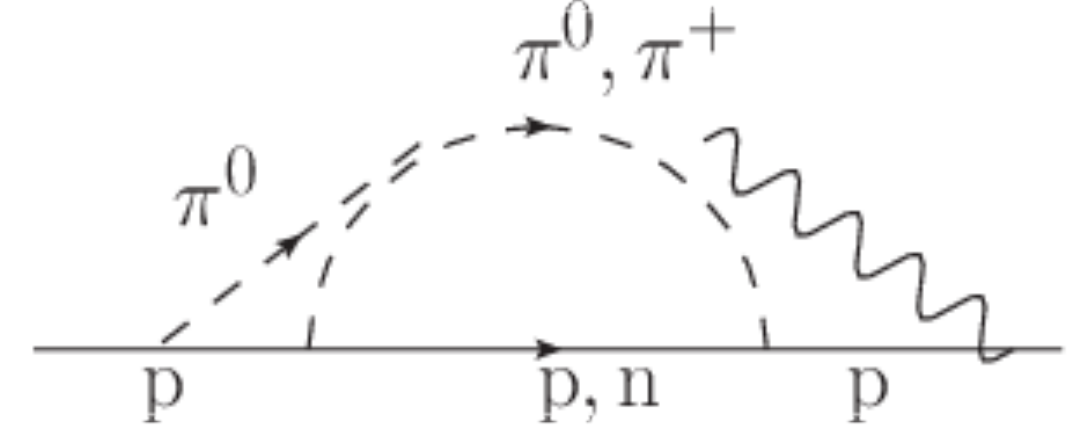}
\includegraphics[width=0.2\textwidth]{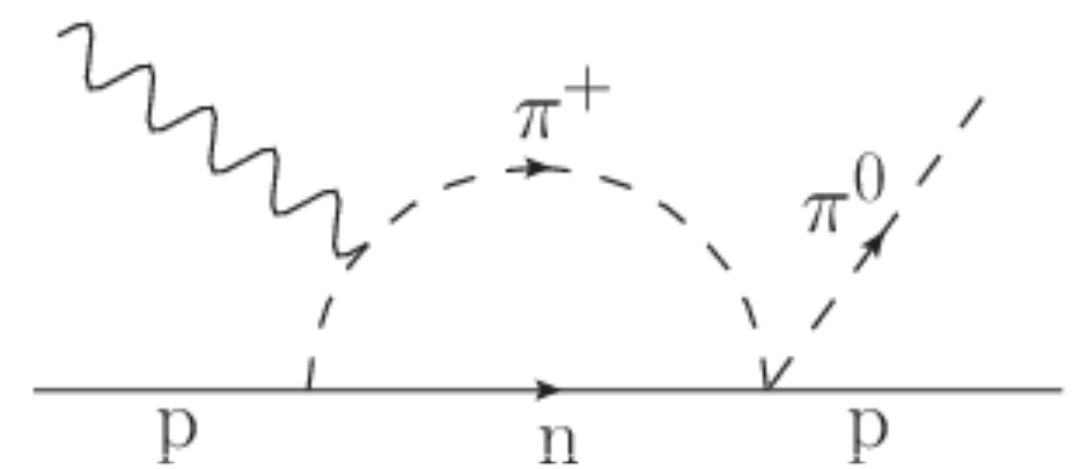}
\includegraphics[width=0.2\textwidth]{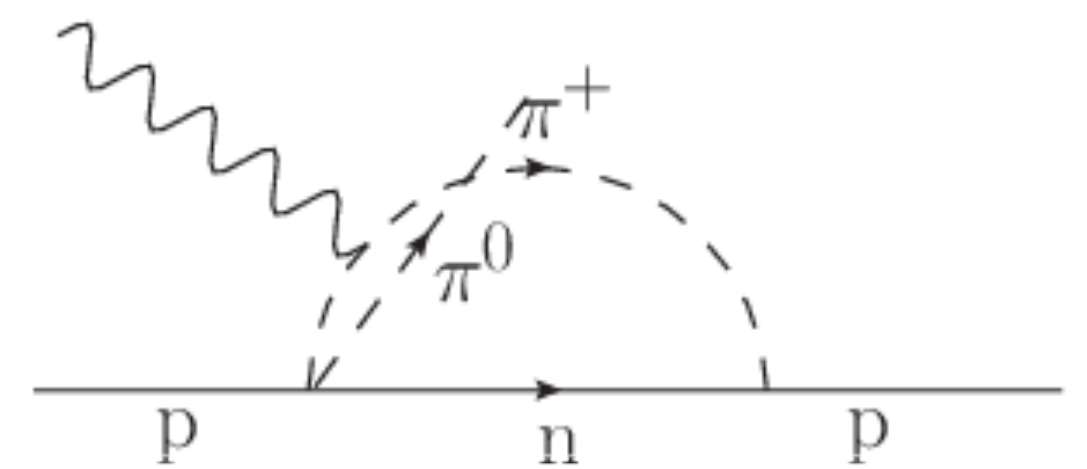}\vspace{5mm}\\
\includegraphics[width=0.2\textwidth]{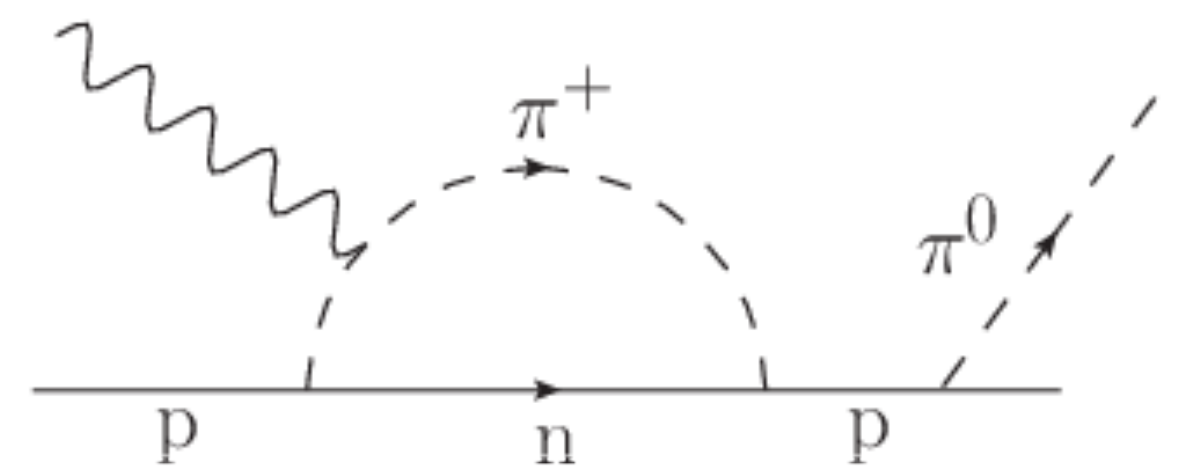}
\includegraphics[width=0.2\textwidth]{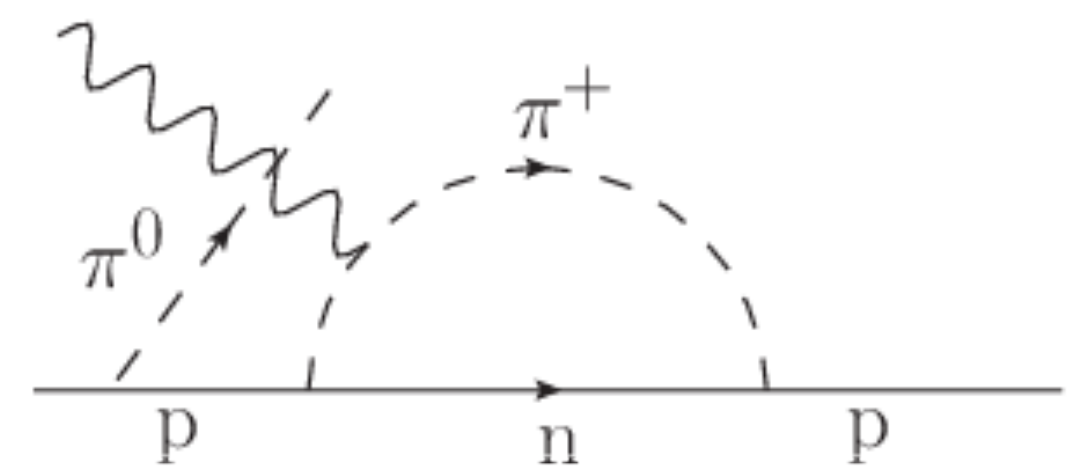}
\includegraphics[width=0.2\textwidth]{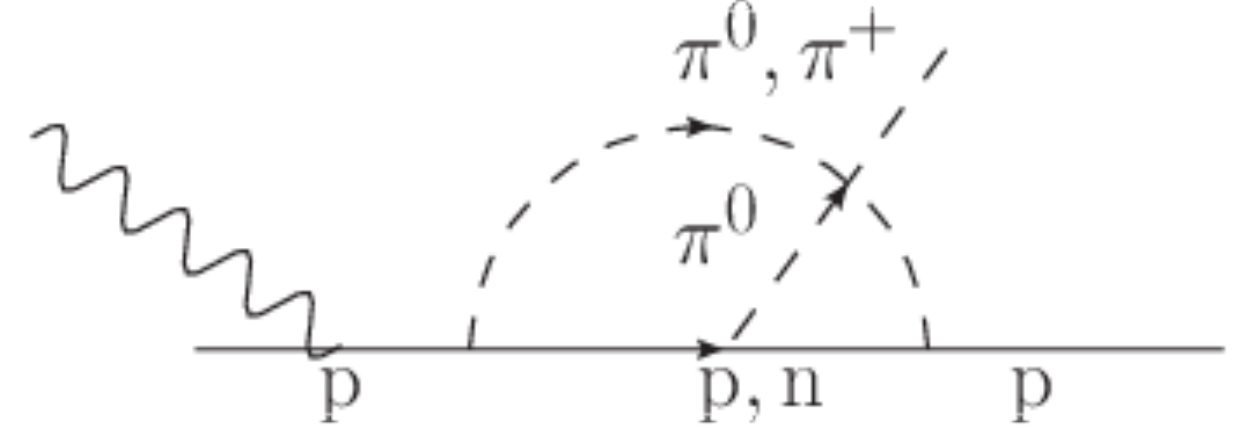}
\includegraphics[width=0.2\textwidth]{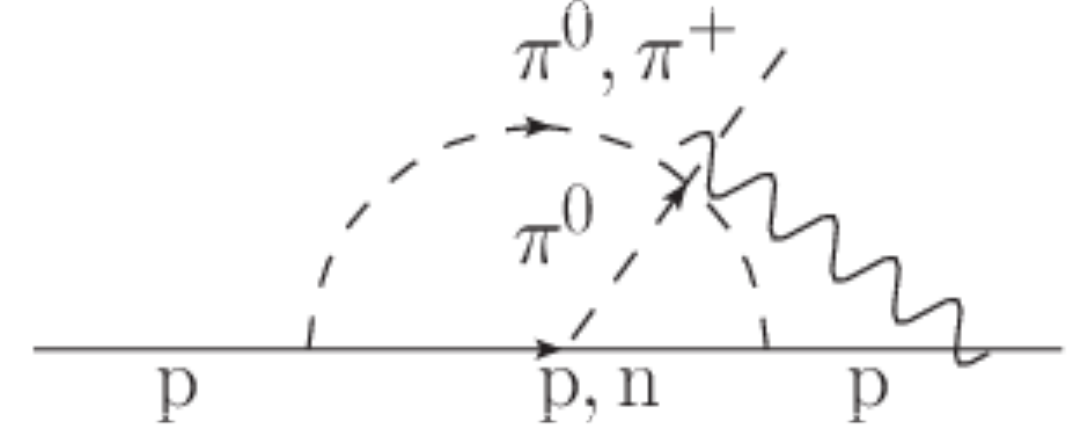}\vspace{5mm}\\
\includegraphics[width=0.2\textwidth]{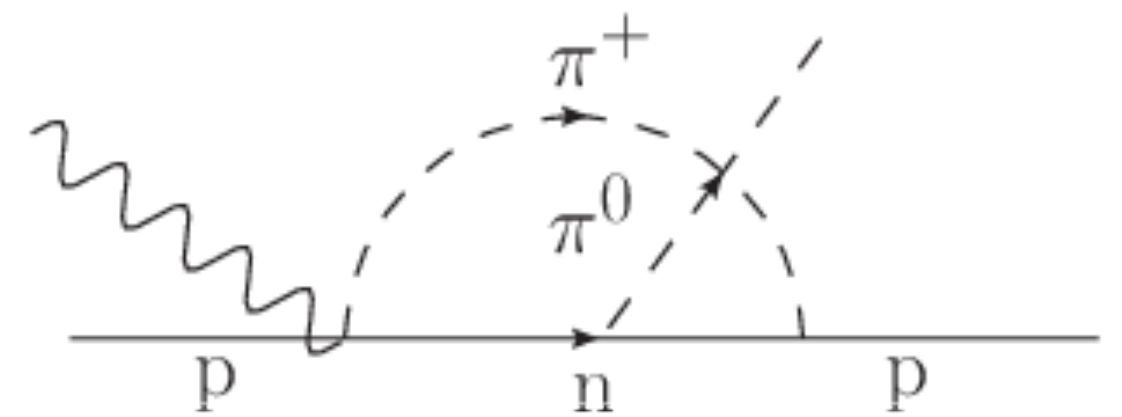}
\includegraphics[width=0.2\textwidth]{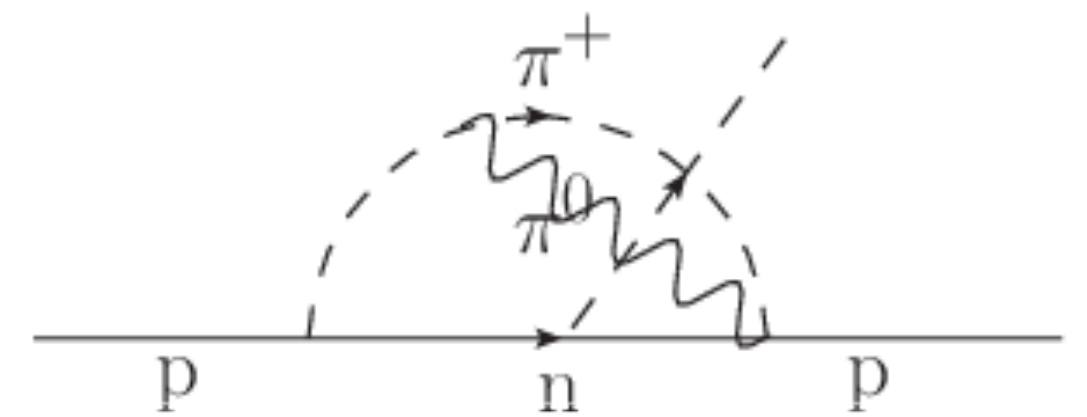}
\includegraphics[width=0.2\textwidth]{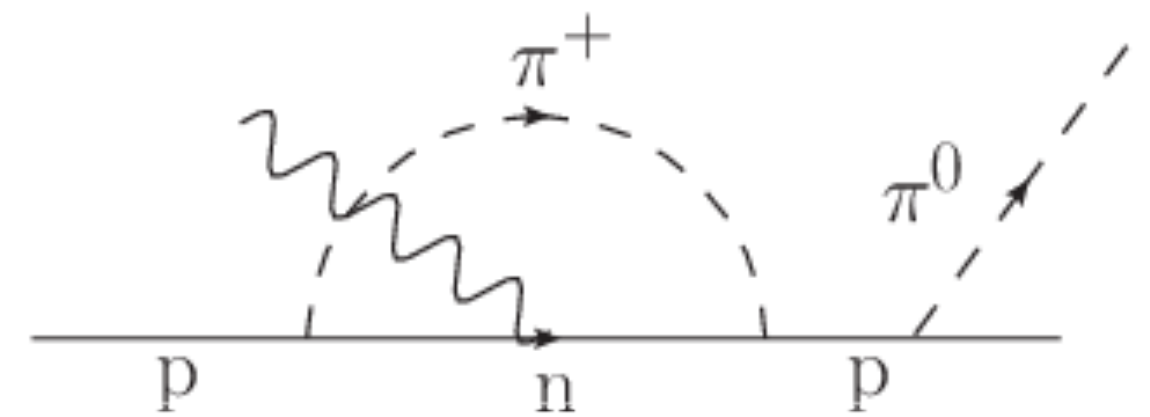}
\includegraphics[width=0.2\textwidth]{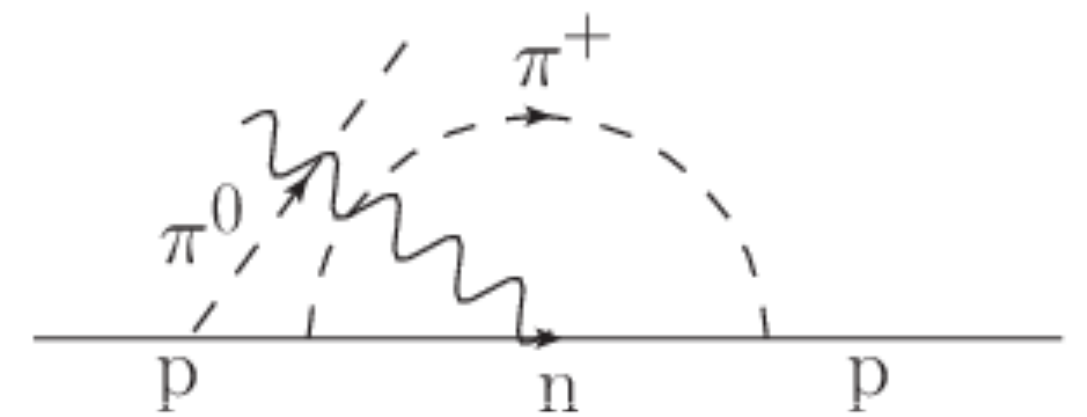}\vspace{5mm}\\
\includegraphics[width=0.2\textwidth]{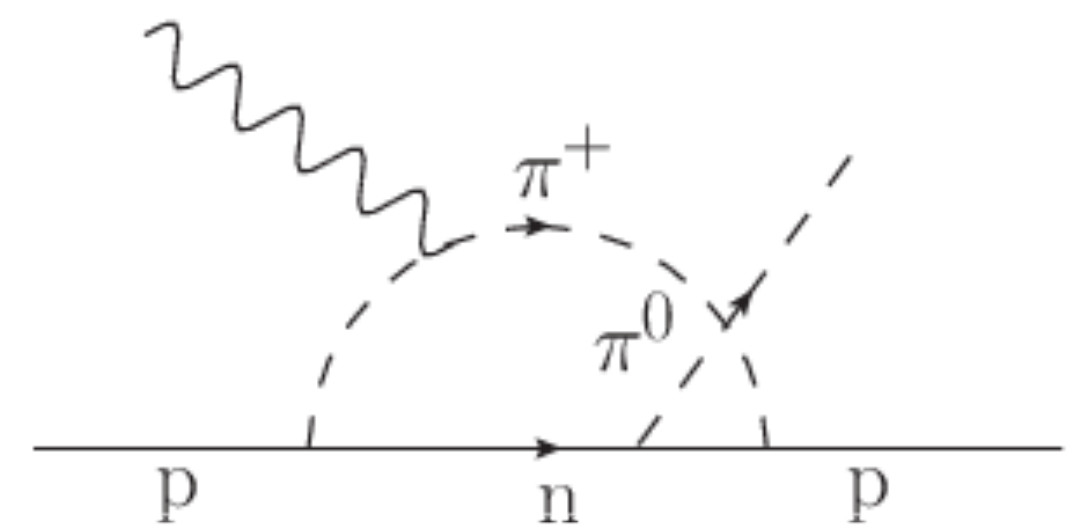}
\includegraphics[width=0.2\textwidth]{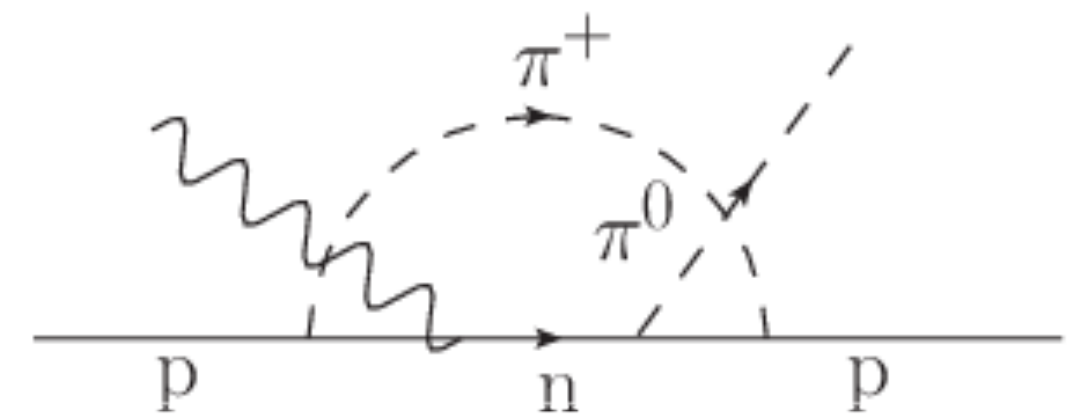}
\includegraphics[width=0.2\textwidth]{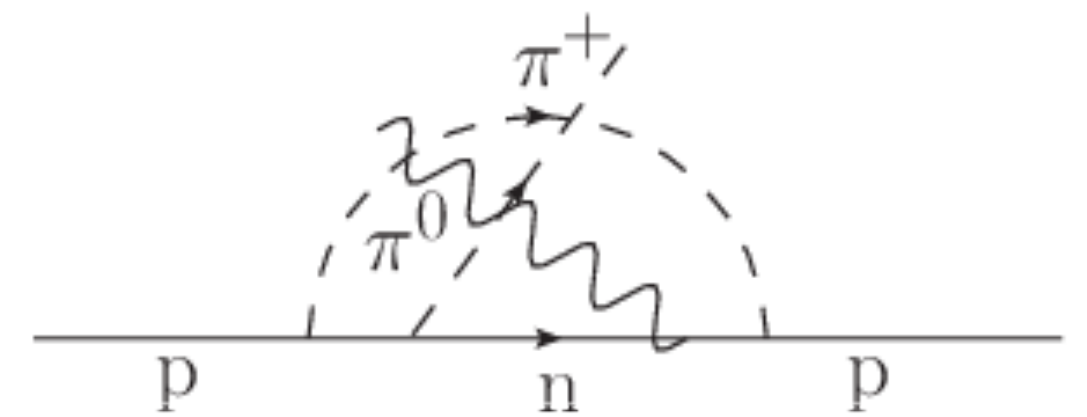}
\end{center}
	\caption{Diagrams with nucleonic virtual states only contributing to the neutral pion photoproduction up to $\mathcal{O}(p^3)$.}
	\label{fDiag12}
\end{figure}

\begin{figure}
\begin{center}
\includegraphics[width=0.4\textwidth]{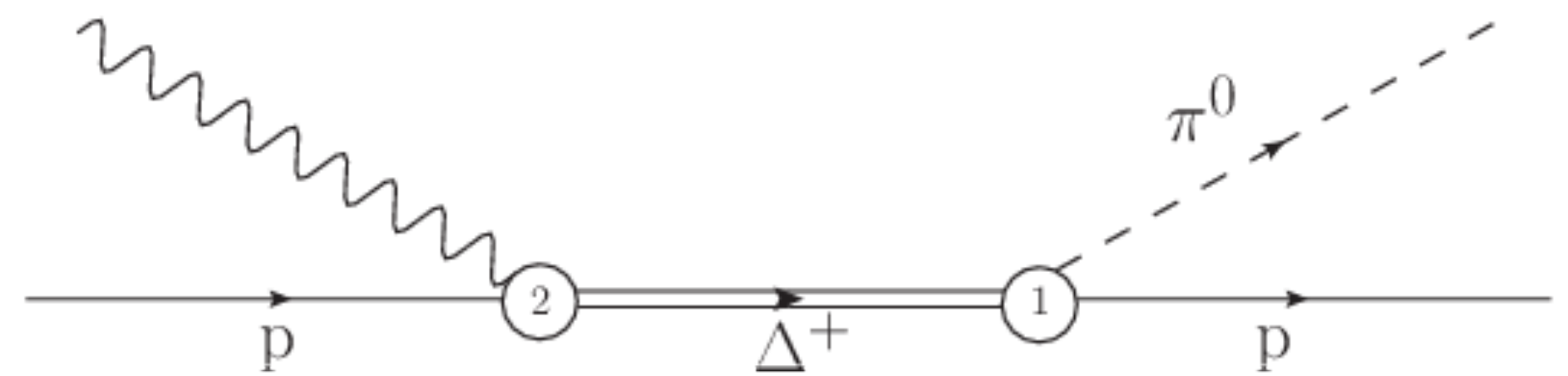}
\includegraphics[width=0.4\textwidth]{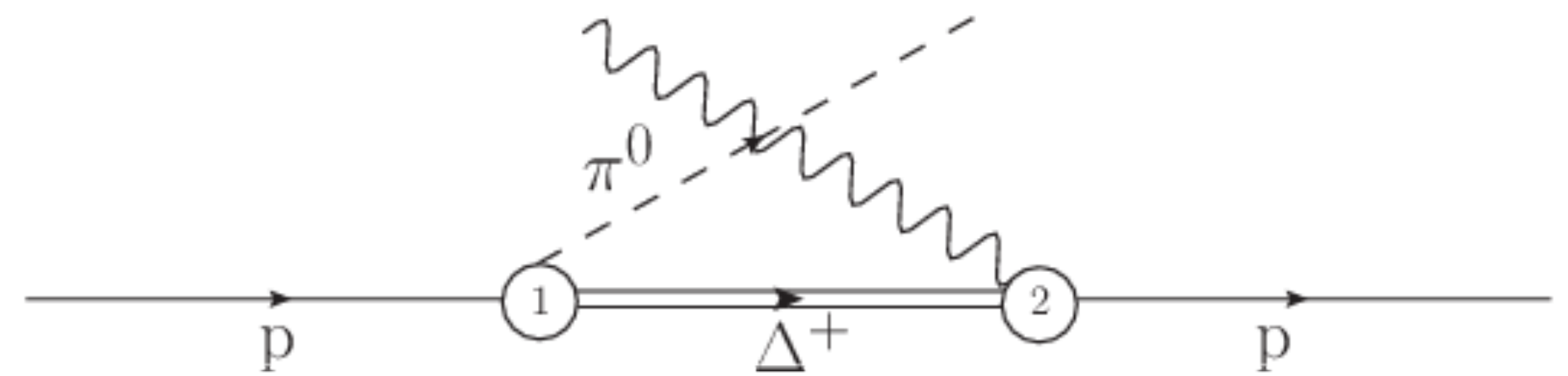}\vspace{5mm}\\
\includegraphics[width=0.4\textwidth]{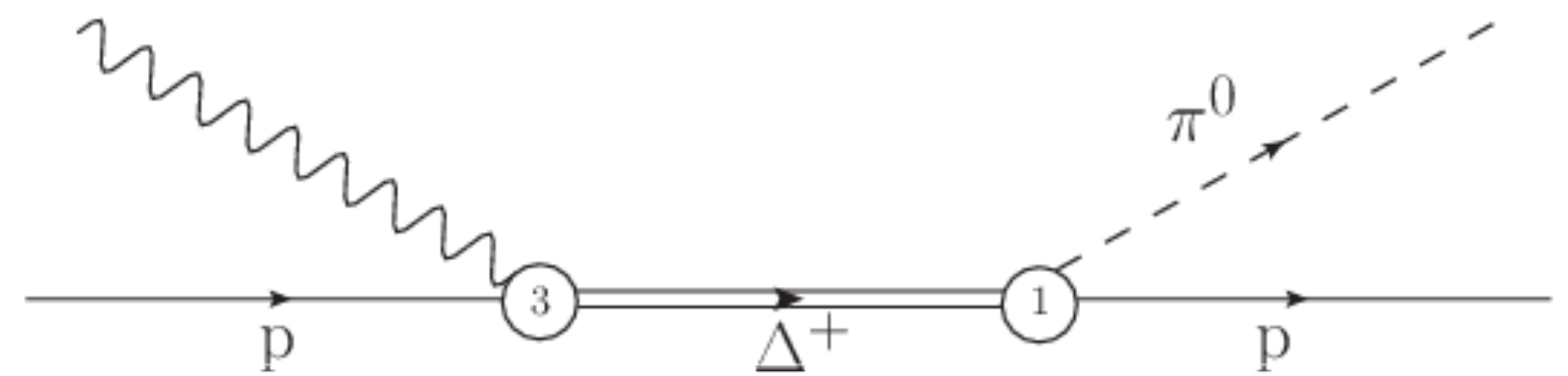}
\includegraphics[width=0.4\textwidth]{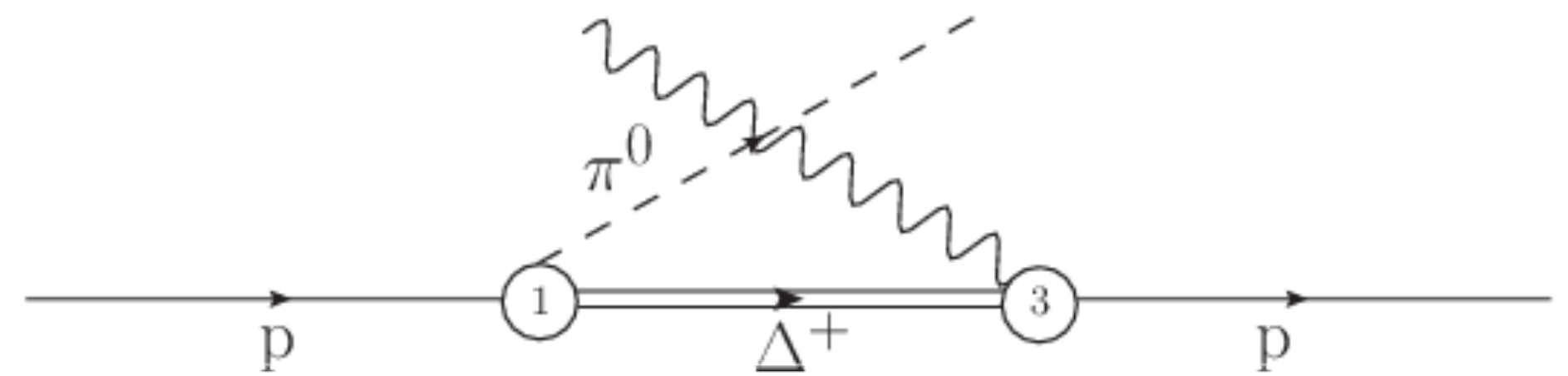}\vspace{5mm}\\
\includegraphics[width=0.3\textwidth]{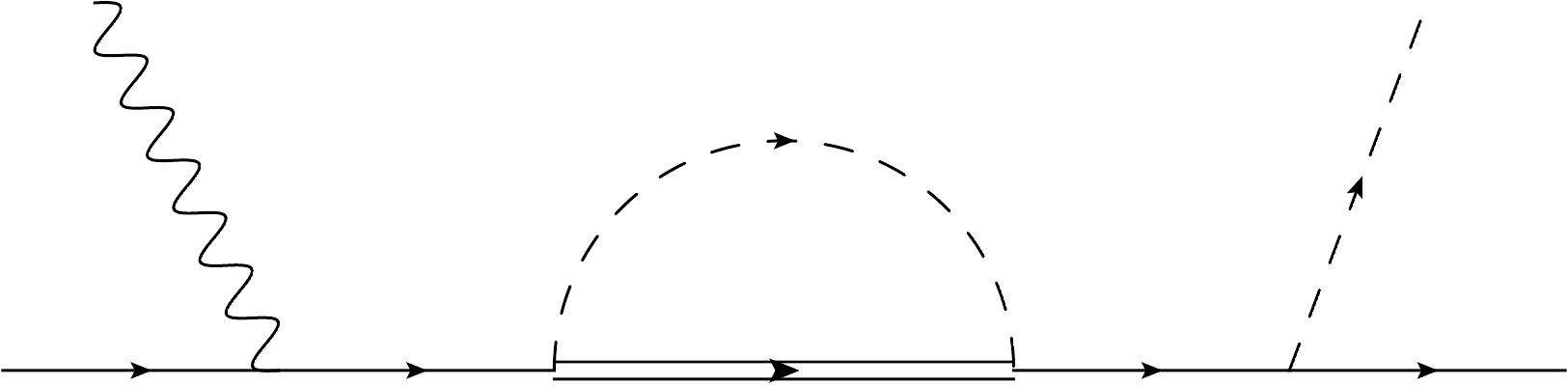}
\includegraphics[width=0.3\textwidth]{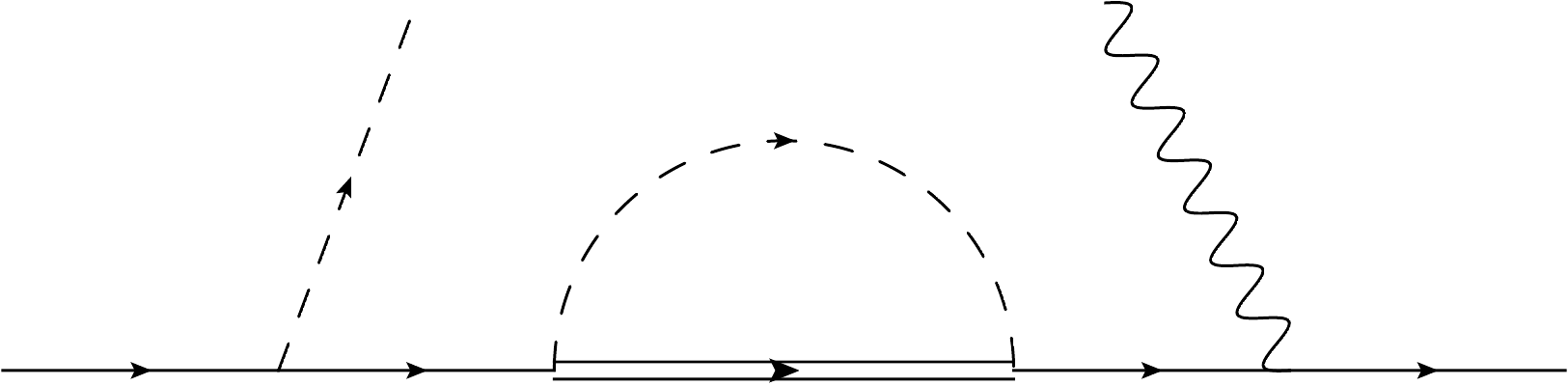}
\includegraphics[width=0.3\textwidth]{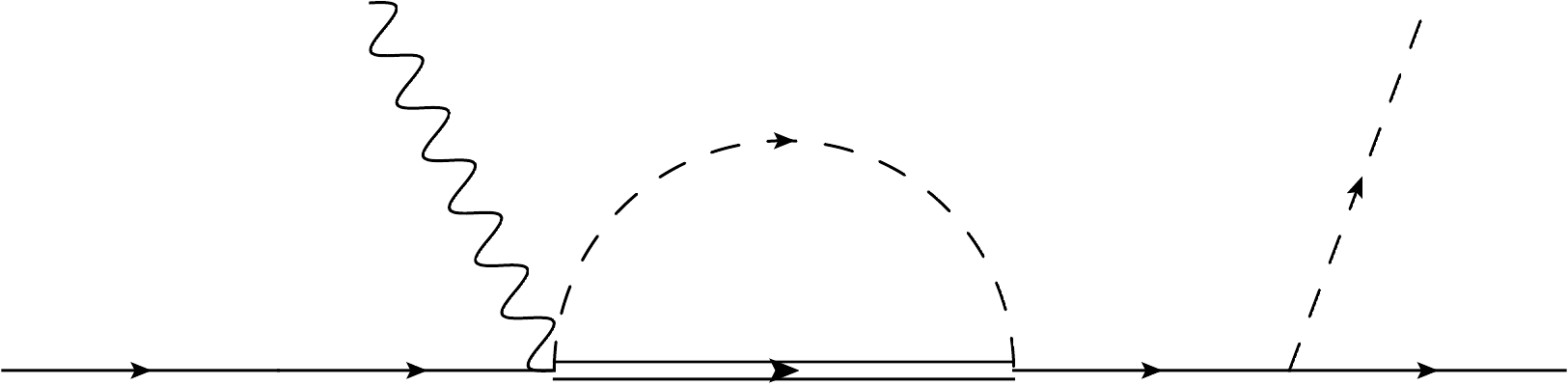}\vspace{5mm}\\
\includegraphics[width=0.3\textwidth]{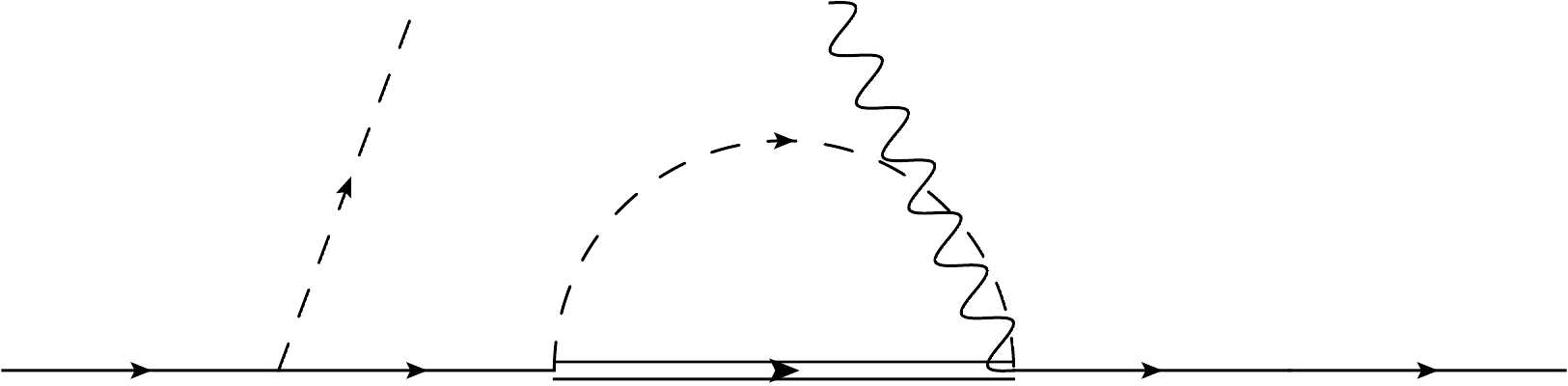}
\includegraphics[width=0.3\textwidth]{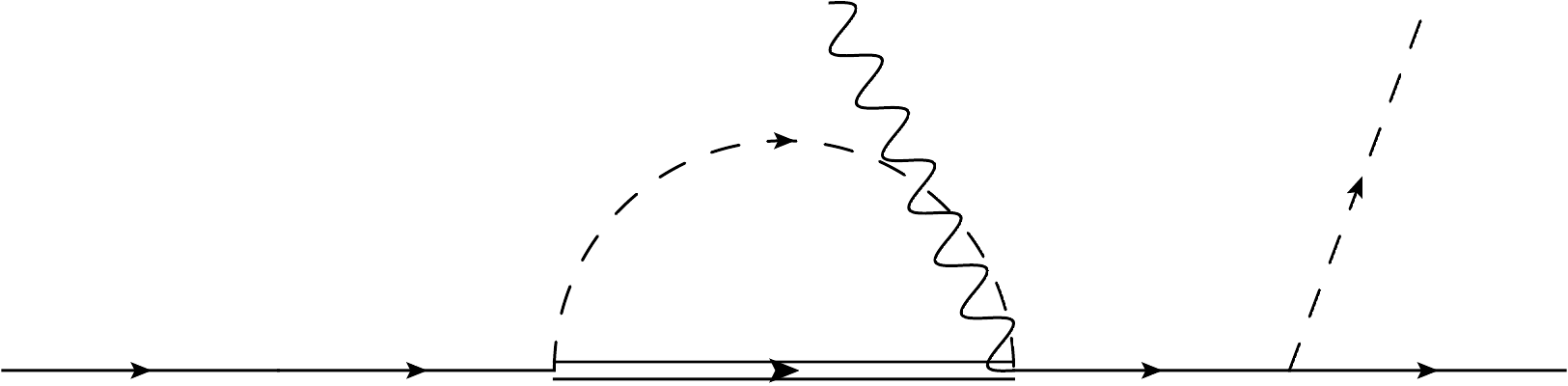}
\includegraphics[width=0.3\textwidth]{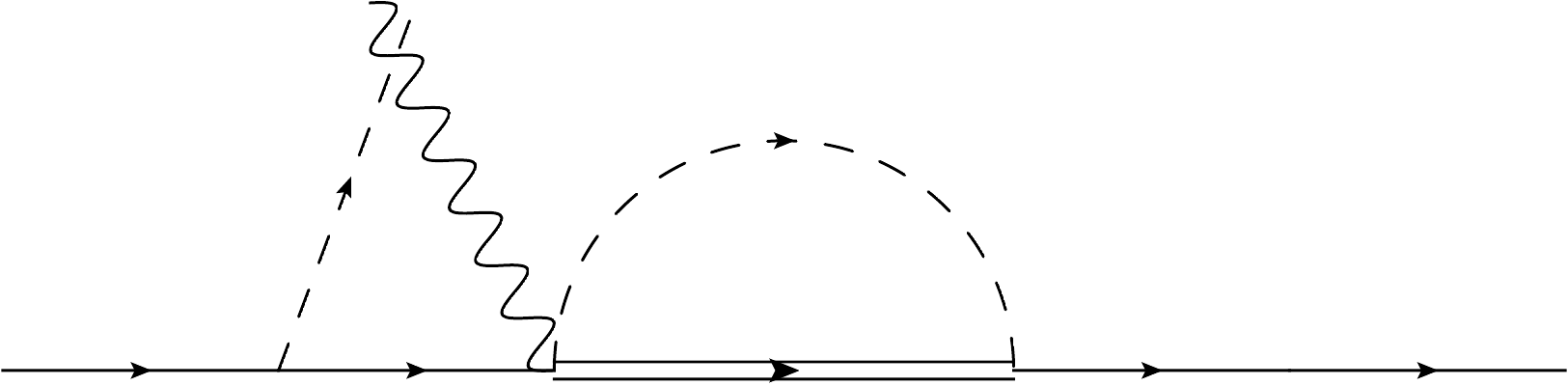}\vspace{5mm}\\
\includegraphics[width=0.3\textwidth]{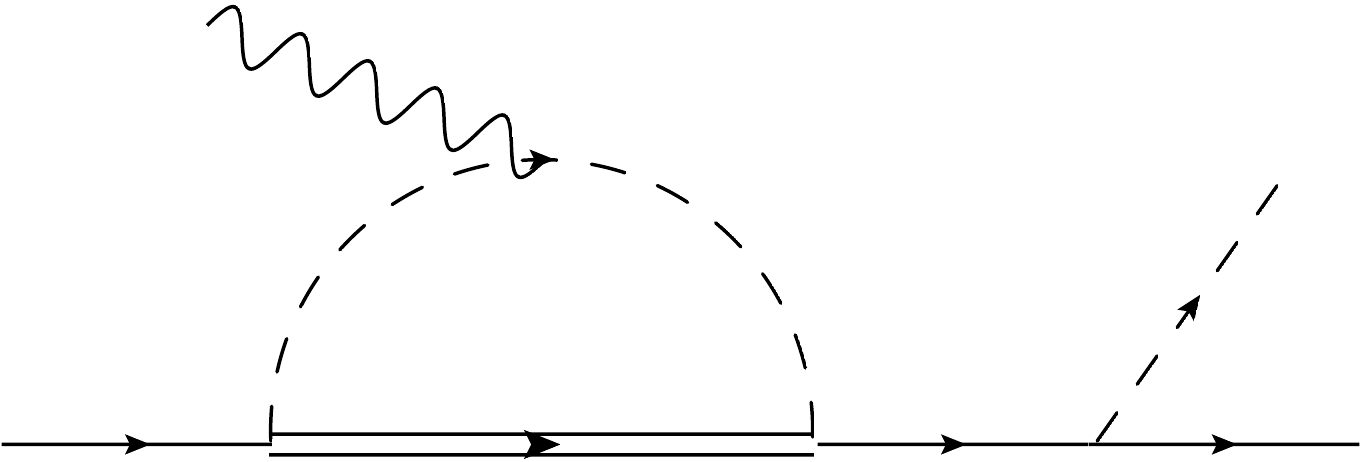}
\includegraphics[width=0.3\textwidth]{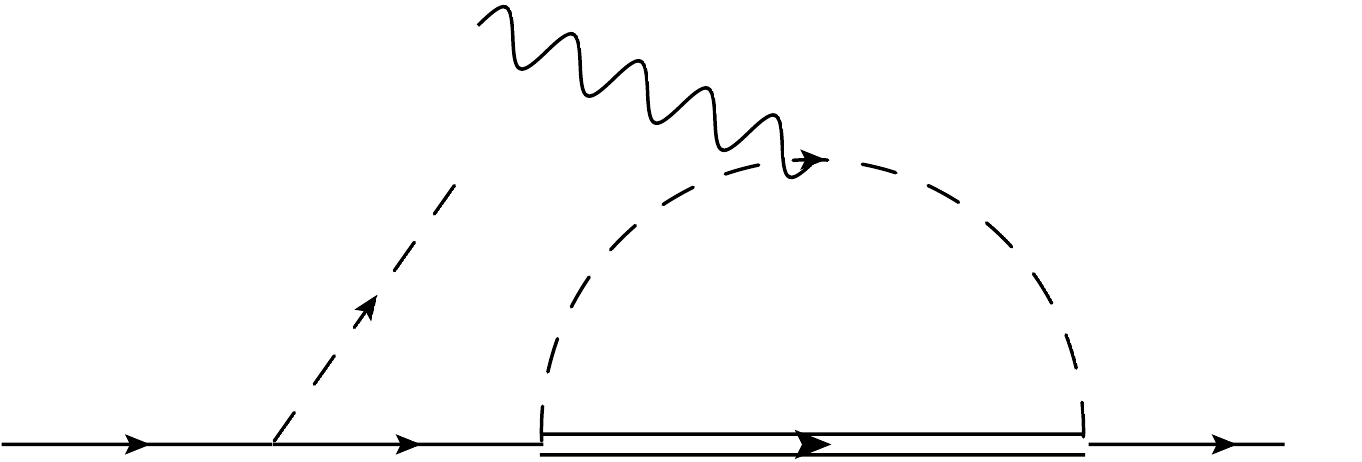}
\includegraphics[width=0.3\textwidth]{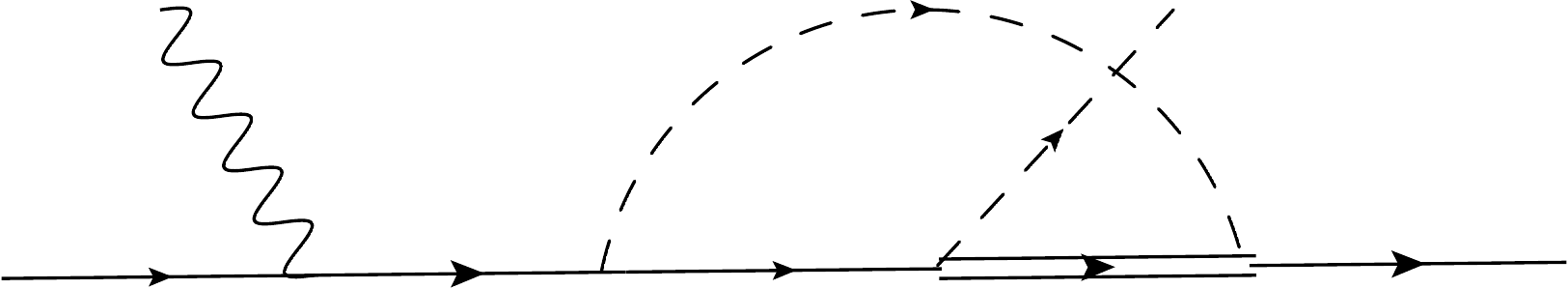}\vspace{5mm}\\
\includegraphics[width=0.3\textwidth]{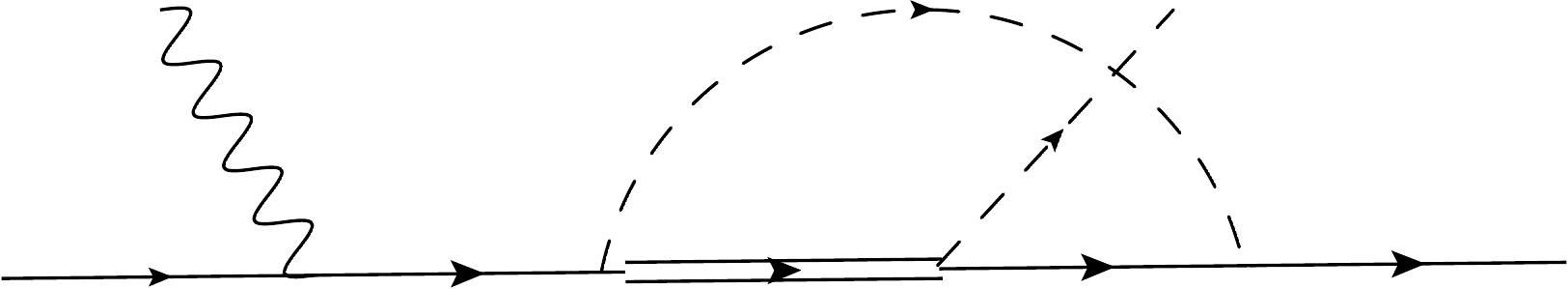}
\includegraphics[width=0.3\textwidth]{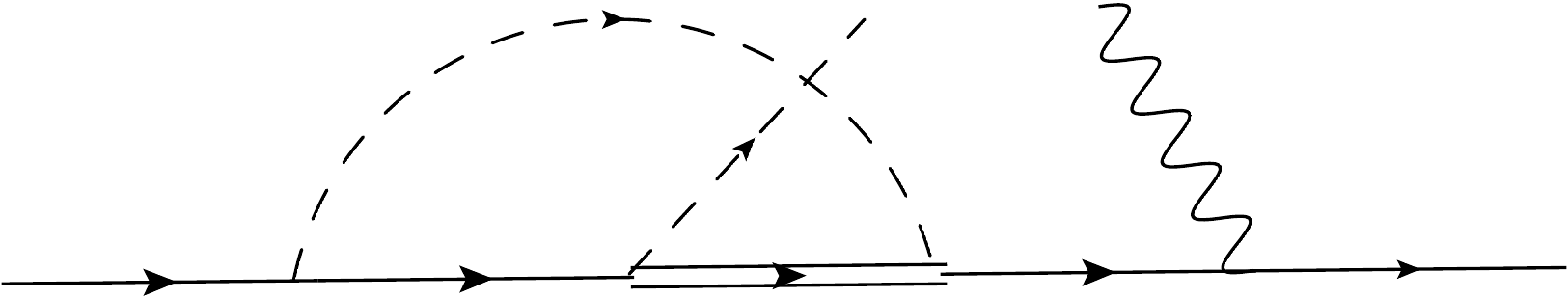}
\includegraphics[width=0.3\textwidth]{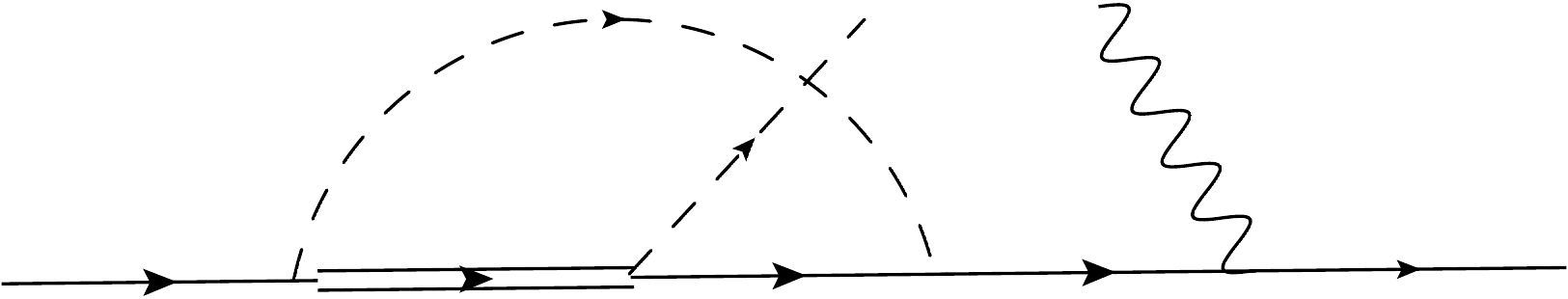}\vspace{5mm}\\
\includegraphics[width=0.3\textwidth]{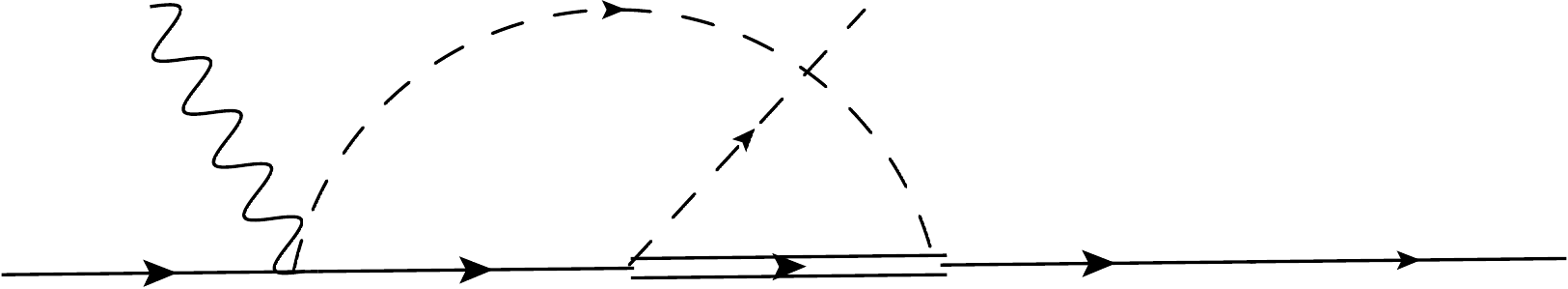}
\includegraphics[width=0.3\textwidth]{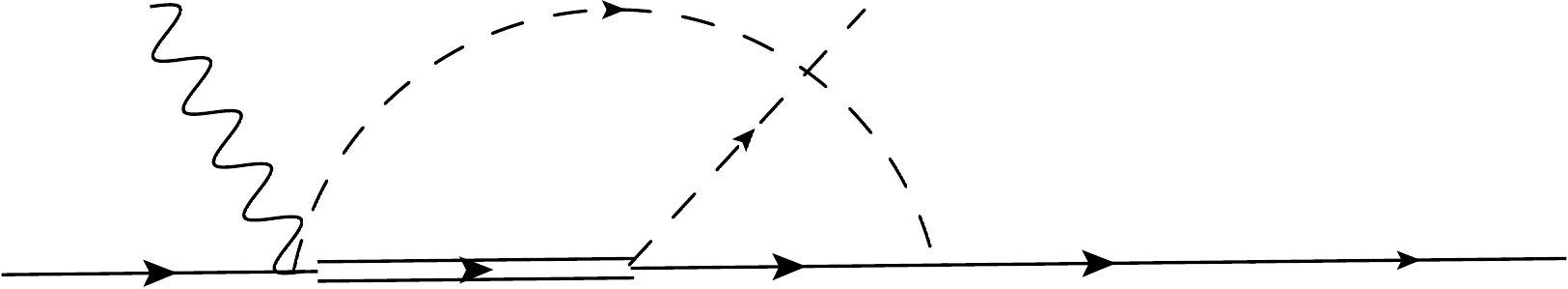}
\includegraphics[width=0.3\textwidth]{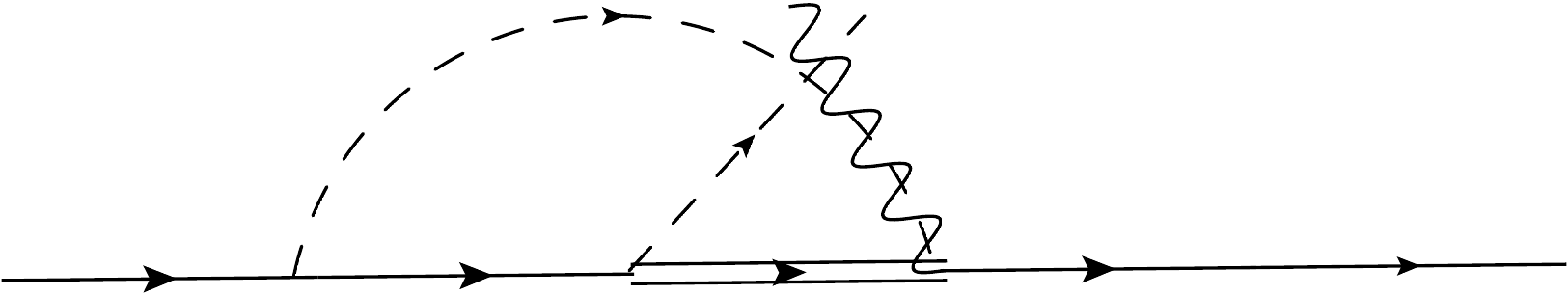}\vspace{5mm}\\
\includegraphics[width=0.3\textwidth]{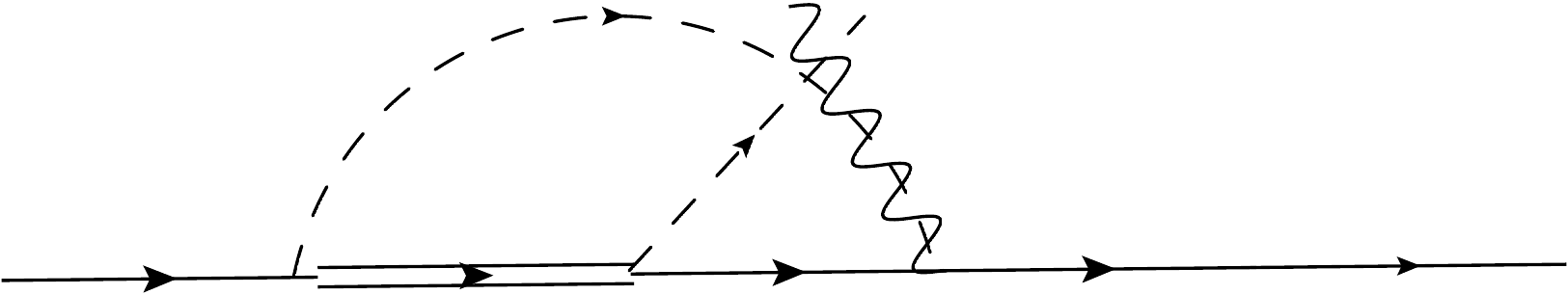}
\includegraphics[width=0.3\textwidth]{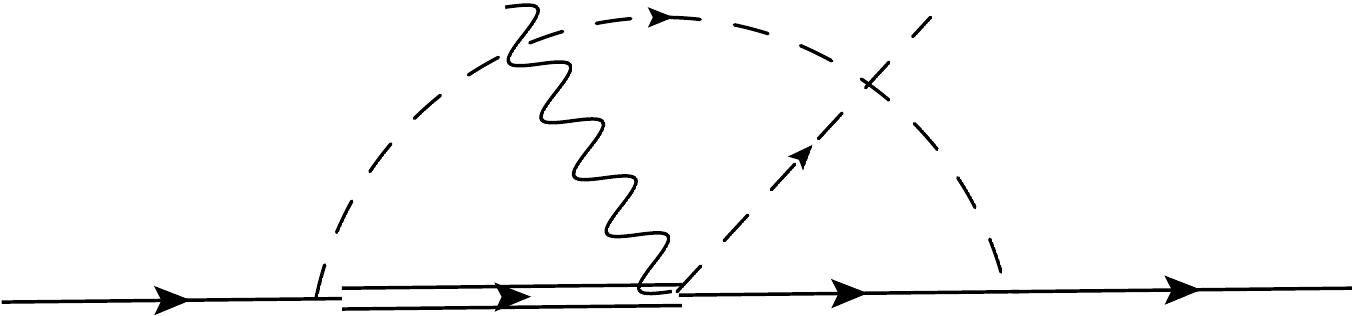}
\includegraphics[width=0.3\textwidth]{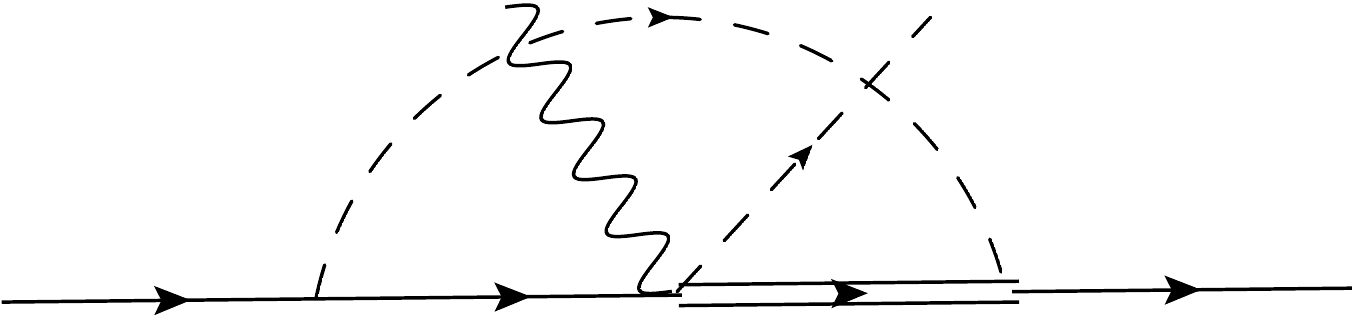}\vspace{5mm}\\
\includegraphics[width=0.3\textwidth]{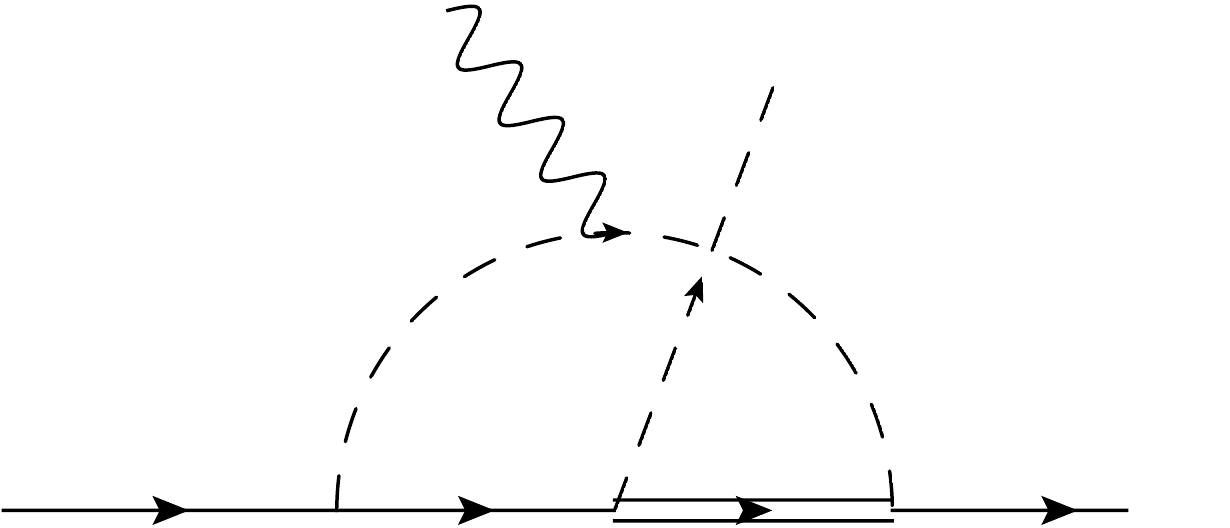}
\includegraphics[width=0.3\textwidth]{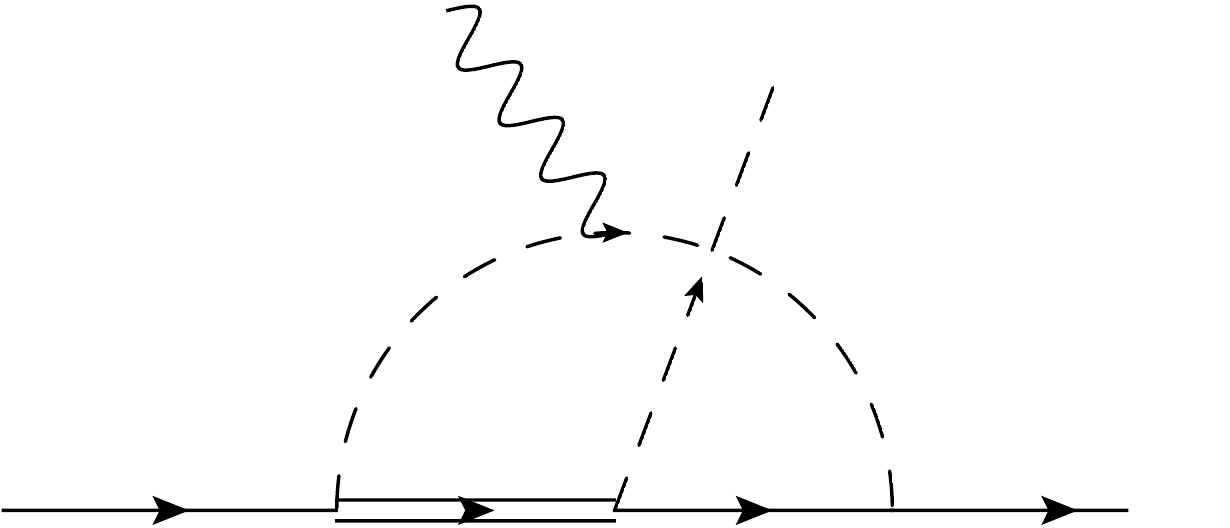}
\includegraphics[width=0.3\textwidth]{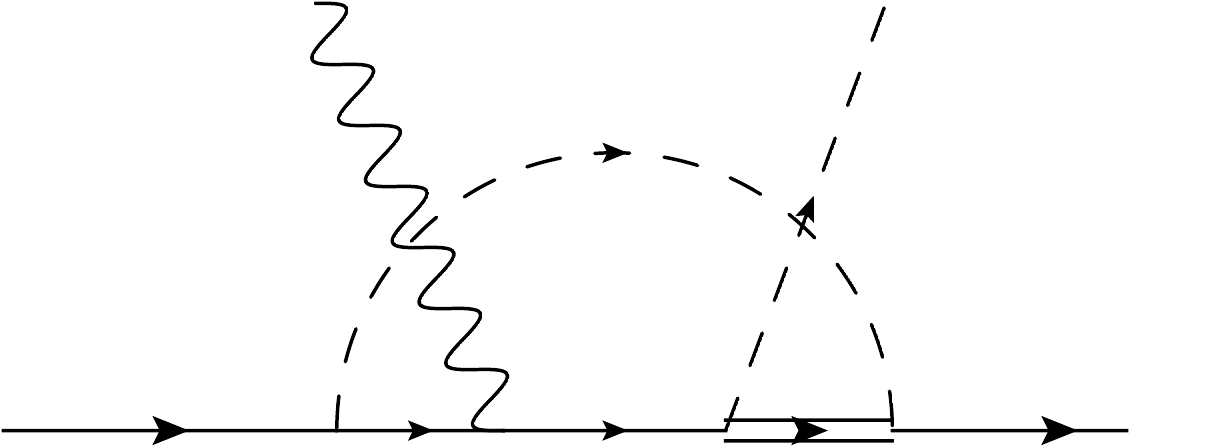}\vspace{5mm}\\
\includegraphics[width=0.3\textwidth]{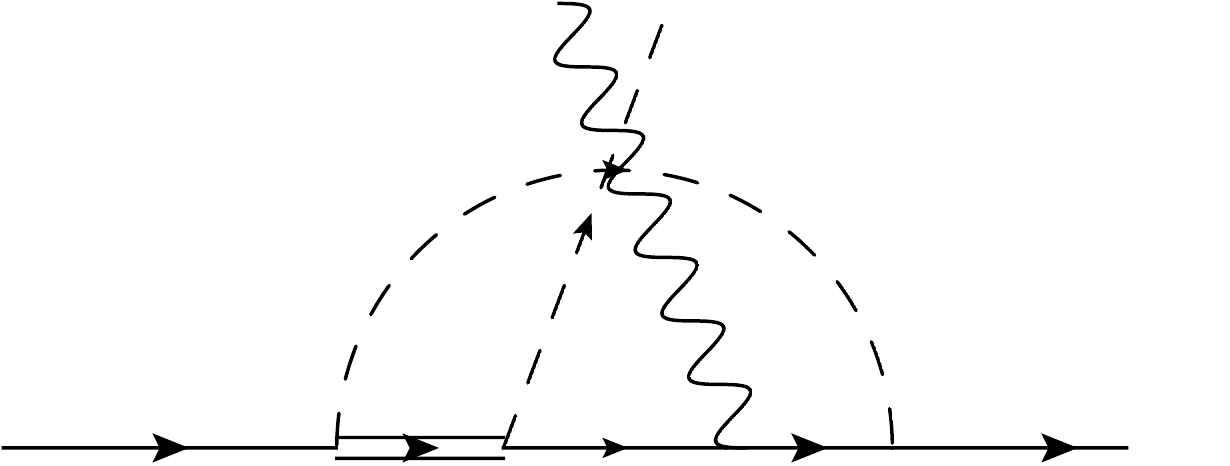}
\includegraphics[width=0.3\textwidth]{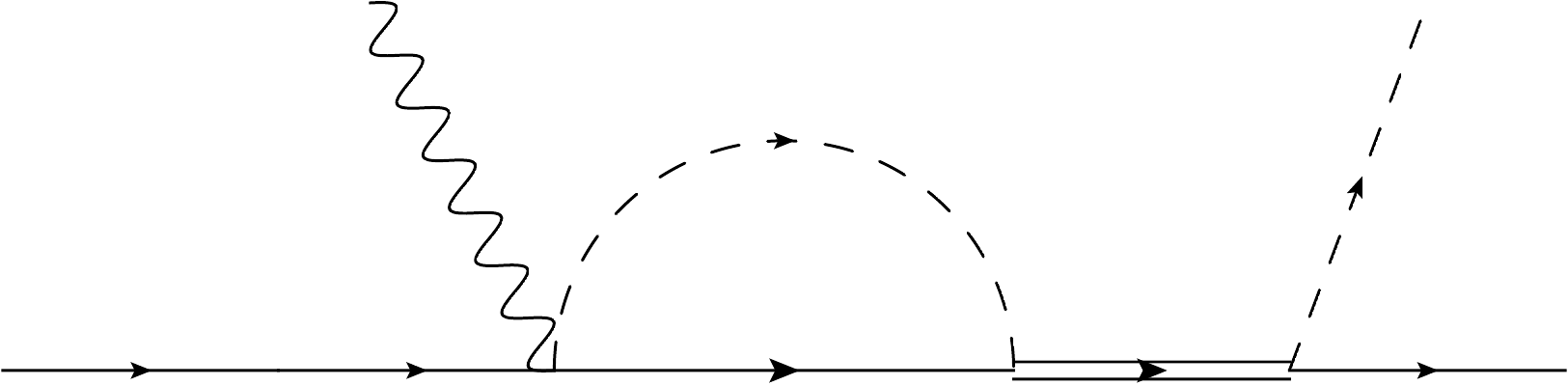}
\includegraphics[width=0.3\textwidth]{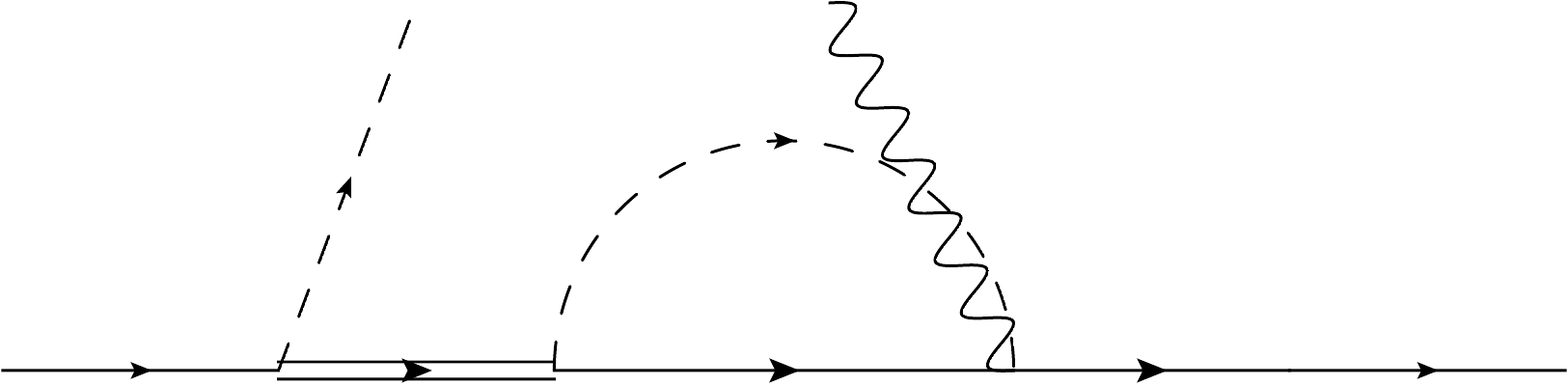}\vspace{5mm}\\
\includegraphics[width=0.3\textwidth]{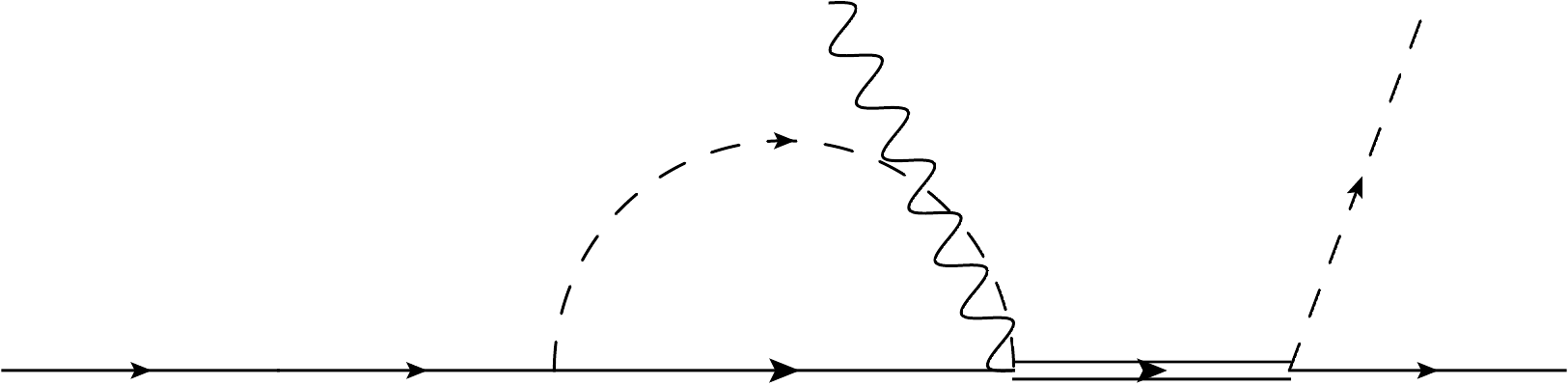}
\includegraphics[width=0.3\textwidth]{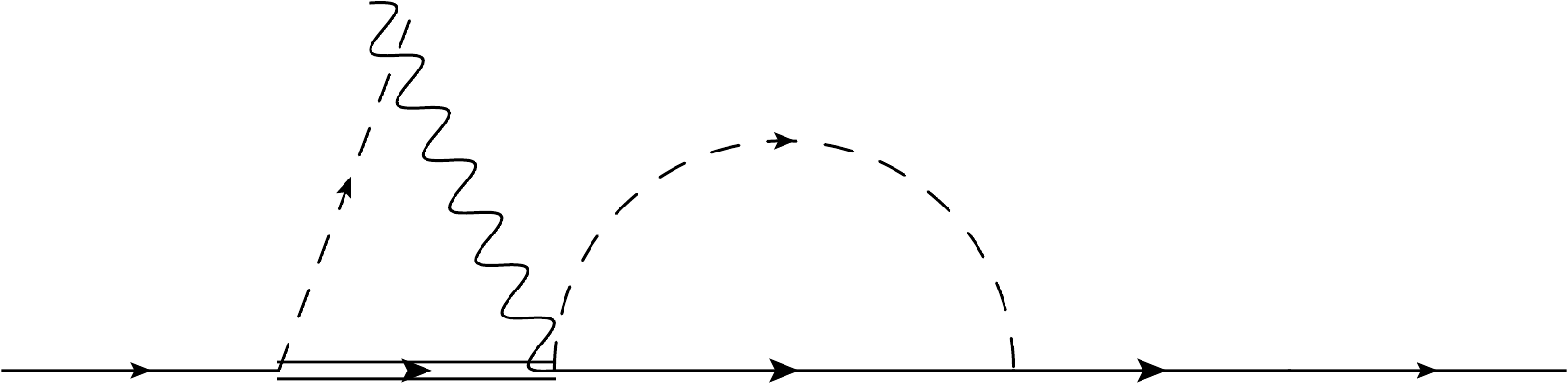}
\includegraphics[width=0.3\textwidth]{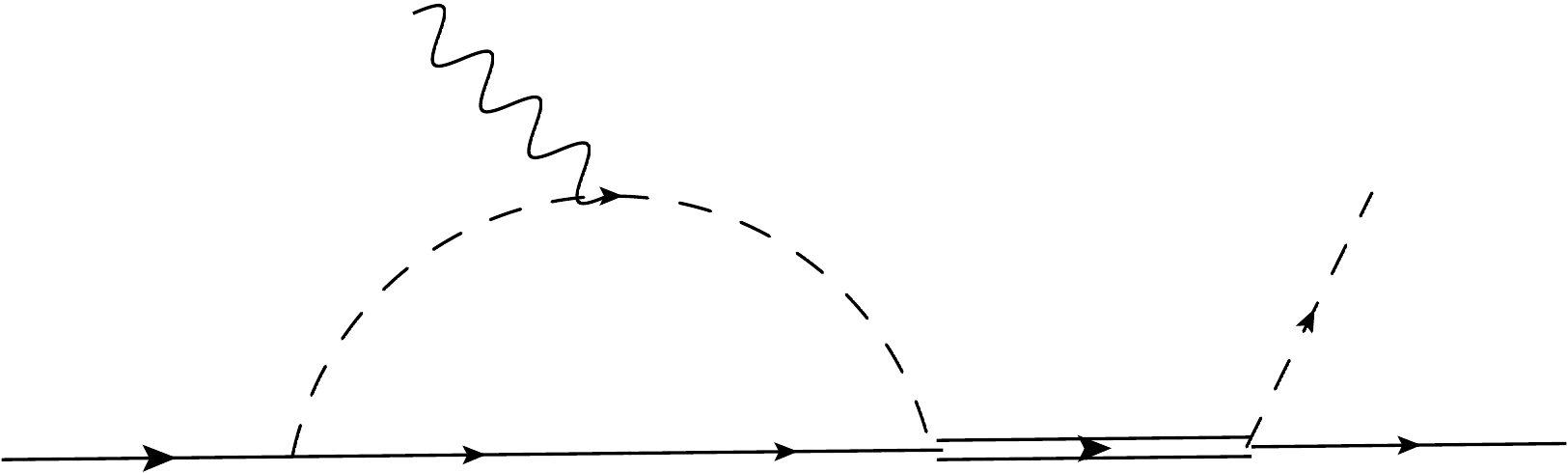}\vspace{5mm}\\
\includegraphics[width=0.3\textwidth]{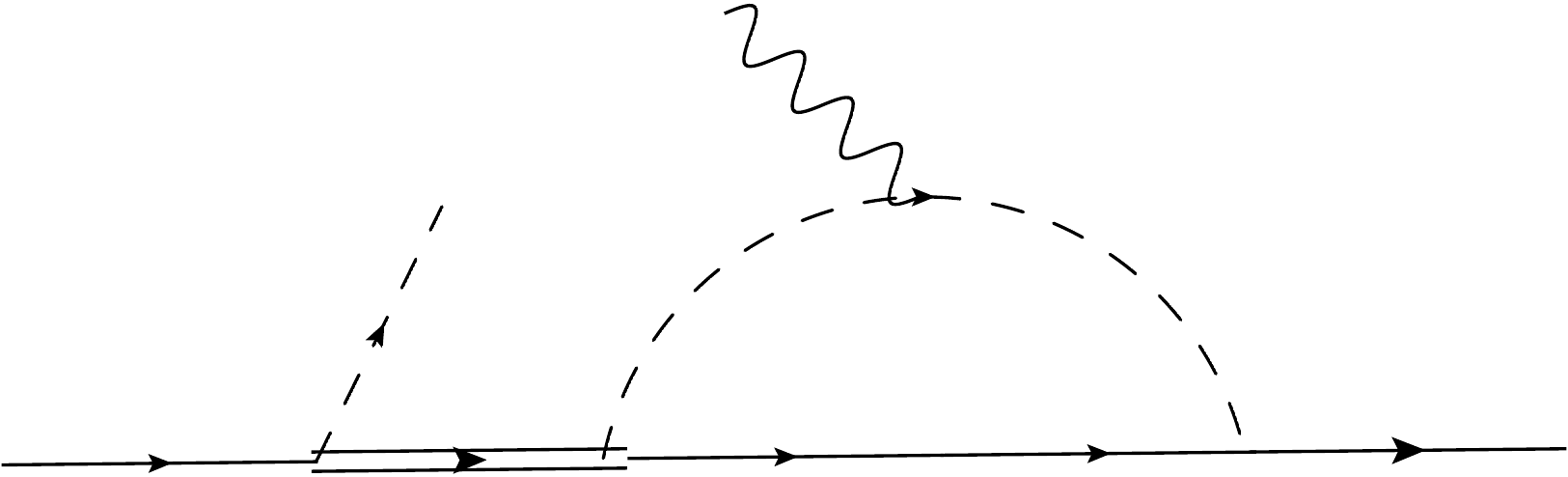}
\includegraphics[width=0.3\textwidth]{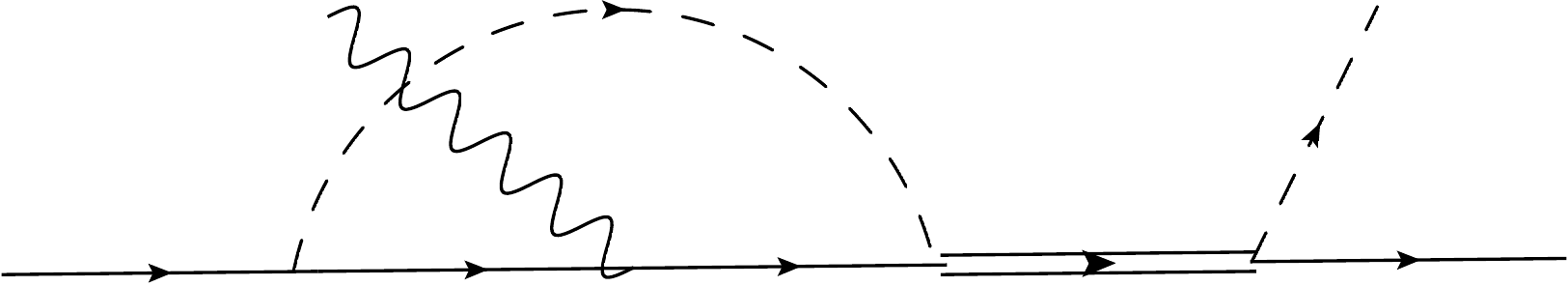}
\includegraphics[width=0.23\textwidth]{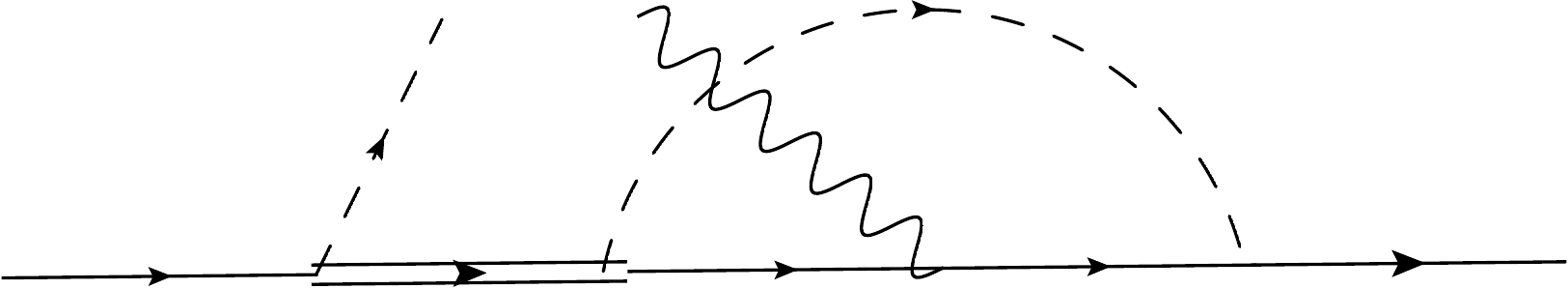}
\end{center}
	\caption{Diagrams with isospin-$3/2$ virtual states contributing to the neutral pion photoproduction up to $\mathcal{O}(p^{7/2})$.}
	\label{fDiag32}
\end{figure}

With the help of the Feynman rules extracted from these Lagrangians, one can now calculate all the amplitudes of the diagrams 
entering this channel at the considered order, shown in Figs.~\ref{fDiag12} and~\ref{fDiag32}. We parameterize the amplitude $\mathcal{M}$ as
\begin{align}
\epsilon_\mu \mathcal{M}^\mu =&\bar u(p')\left(V_Nq\cdot\epsilon\gamma_5+V_Kq\cdot\epsilon\slashed{k}\gamma_5+V_E\slashed{\epsilon}\gamma_5+V_{EK}\slashed{\epsilon}\slashed{k}\gamma_5\right)u(p),
\end{align}
where $V_N$, $V_K$, $V_E$ and $V_{EK}$ are structure functions of the photon energy and the scattering angle, 
and where $\epsilon_\mu$ is the photon-polarization 4-vector, $k_\mu$ its 4-momentum and $q_\mu$ the momentum of the outgoing $\pi^0$. The Dirac spinors $u(p)$ and $\bar u (p')$ are those of the nucleon in the initial and final states, respectively. Another common representation is current conserving by definition and has the form~\cite{Drechsel:1992pn}
\begin{align}
\nonumber \epsilon_\mu \mathcal{M}^\mu = &\epsilon_\mu \bar u(p')\left(
	\sum_{i=1}^4A_iM_i^\mu\right)u(p),
\end{align}
with
\begin{align}
\nonumber \epsilon\cdot M_1 =& \mathrm{i} \slashed{k}\slashed{e}\gamma_5,\\
 \nonumber \epsilon\cdot M_2 =&\mathrm{i}(p'\cdot\epsilon k\cdot q - q\cdot\epsilon k\cdot(p+p'))\gamma_5,\\
\nonumber \epsilon\cdot M_3 =&\mathrm{i}(\slashed{e} k\cdot q- \slashed{k} q\cdot\epsilon)\gamma_5,\\
 \nonumber \epsilon\cdot M_4 =&\mathrm{i}(\slashed{e} k \cdot(p+p')-\slashed{k} p'\cdot \epsilon - 2m \slashed{k}\slashed{e})\gamma_5.
\end{align}
The conversion between parameterizations is straightforward:
\begin{align}
\nonumber A_1 = & 
\mathrm{i}\left(V_{EK}-\frac{m}{k\cdot p}\left(V_E+k\cdot q V_K\right)\right),\\
\nonumber A_2 = &
\mathrm{i}\frac{V_N}{2 k\cdot p},\\
\nonumber A_3 = &
\mathrm{i}\left(V_K\left(1-\frac{k\cdot q}{2k\cdot p}\right)-\frac{V_E}{2k\cdot p}\right),\\
\nonumber A_4 = &
-\frac{\mathrm{i}}{2k\cdot p}\left(V_E+k\cdot qV_K\right).
\end{align}
Finally, for the calculation of multipoles it is useful to write the expressions in terms of the Chew--Goldberger--Low--Nambu (CGLM) amplitudes~\cite{Chew:1957tf}, 
\begin{align}
\nonumber \epsilon_\mu \mathcal{M}^\mu = \frac{4\pi W}{m}\chi_f^\dagger\mathcal{F}\chi_i,
\end{align}
where $\chi_i$ and $\chi_f$ are the initial and final state Pauli spinors, respectively, and $W$ the center of mass energy. For real photons, the amplitude $\mathcal{F}$ may be written as
\begin{align}
\nonumber \mathcal{F} = \mathrm{i}\vec\tau\cdot\vec\epsilon\mathcal{F}_1
+  \vec\tau\cdot\hat q\vec\tau\cdot\hat k\times\vec\epsilon\mathcal{F}_2
+  \mathrm{i}\vec\tau\cdot\hat k \hat q\cdot \epsilon\mathcal{F}_3
+  \mathrm{i}\vec\tau\cdot\hat q \hat q\cdot \epsilon\mathcal{F}_4.
\end{align}
The conversion between parameterizations is given by
\begin{align}\nonumber
\mathcal{F}_1 =&\frac{\sqrt{(E_i+m)(E_f+m)}}{8\pi W}\Bigg[
-\left(k_0 +\frac{k_0^2}{E_i+m}\right)A_1-k\cdot q A_3\\
\nonumber&+\left(-k_0^2+2k_0m+\frac{2k_0^2m}{E_i+m}-k_0(E_i+E_f)-k_0|\vec q|\cos\theta\right)A_4\Bigg],
\end{align}
\begin{align}
 \nonumber\mathcal{F}_2 =&\frac{\sqrt{(E_i+m)(E_f+m)}}{8\pi W}|\vec q|\Bigg[\left(\frac{k_0}{E_f+m} +\frac{k_0^2}{(E_i+m)(E_f+m)}\right)A_1\\
\nonumber&
-\frac{k_0k\cdot q}{(E_i+m)(E_f+m)}A_3\\
\nonumber&-\left(k_0\frac{k_0^2+2k_0m+k_0(E_i+E_f)+k_0|\vec q|\cos\theta}{(E_i+m)(E_f+m)}
+\frac{2k_0m}{E_f+m}\right)A_4\Bigg],
\end{align}
\begin{align}
\nonumber\mathcal{F}_3 =& \frac{\sqrt{(E_i+m)(E_f+m)}}{8\pi W}|\vec q|\Bigg[
-k_0^2\frac{E_i+E_f + k_0 + q_0}{E_i+m}A_2\\
\nonumber&+\left(k_0+\frac{k_0^2}{E_i+m}\right)  (A_4 - A_3)
\Bigg],
\end{align}
\begin{align}
\nonumber\mathcal{F}_4 =&\frac{\sqrt{(E_i+m)(E_f+m)}}{8\pi W}|\vec q|^2\Bigg[
\left(k_0\frac{k_0+E_i+E_f+q_0}{E_f+m}\right)A_2\\
\nonumber &+\left(\frac{k_0}{E_f+m}+\frac{k_0^2}{(E_i+m)(E_f+m)}\right) (A_4 - A_3)
\Bigg].
\end{align}

The clear ordering of the mesonic ChPT is spoiled by the inclusion of baryons: 
power-counting breaking terms appear in those diagrams that have baryons inside of a loop. When integrating over the loop 
momenta, in addition to the usual divergences of dimensional regularization, there are now also terms that belong to a lower 
chiral order than the nominal order of the diagram. A scheme which has proven to be straightforward and 
effectively renormalizes both these issues is the EOMS regularization scheme~\cite{Gegelia:1999gf,Fuchs:2003qc}. The expressions of 
the divergences and power-counting breaking terms are fully analytical. Therefore they are absorbed into the low-energy constants of 
the corresponding order. This scheme has been broadly studied in~\cite{Fuchs:2003ir,Lehnhart:2004vi,Schindler:2006it,Schindler:2006ha,Geng:2008mf,Geng:2009ik,MartinCamalich:2010fp,Alarcon:2011zs,Ledwig:2011cx,Chen:2012nx,Scherer:2012xha,Alvarez-Ruso:2013fza,Ledwig:2014rfa}. The specific subtraction we perform is the modified 
minimal subtraction $\widetilde{MS}$~\cite{Scherer:2012xha}. For that purpose, all terms proportional to
\begin{align*}
L=\frac{2}{\epsilon}+\log(4\pi)-\gamma_E+1
\end{align*}
are subtracted, with $\epsilon=4-\text{dim}$. Then, after making an expansion of the amplitudes\footnote{The chosen small parameters were $m_\pi$, $\nu=(s-u)/(4 m)$ with $s$ and $u$ the Mandelstam variables of $\mathcal{O}(p^1)$, and the Mandelstam variable $t$ of order $\mathcal{O}(p^2)$ as in ~\cite{Alarcon:2012kn}.},
we have removed the power-counting breaking terms. The analytical expression obtained for the power-counting breaking terms in the isospin-$1/2$ sector reads
\begin{align*}
\frac{\mathrm{i} e g_A^3 m}{32F^3 \pi^2}\left[
\left(4\nu - 3\frac{m_\pi^2}{\nu}\right)\slashed{\epsilon}\gamma_5
+\left(3 - 3\frac{m_\pi^2}{\nu^2}\right)\slashed{\epsilon}\slashed{k}\gamma_5
+ \frac{1}{\nu}q\cdot\epsilon\slashed{k}\gamma_5
-\frac{2m}{\nu}q\cdot\epsilon\gamma_5
\right].
\end{align*}
The expressions of the additional power-counting breaking terms coming from the introduction of the $\Delta$ loops are larger and thus not shown here.

At the considered order, the wave-function renormalization has to be taken into account for the external proton legs  of the $\mathcal{O}(p^1)$ tree diagrams, as the correction amounts to multiplying this tree-level amplitude by the residuum $Z_p$, a term of $\mathcal{O}(p^2)$. All the corrections to higher-order amplitudes or to the other external legs would be at least of $\mathcal{O}(p^4)$. The analytical expression for this correction factor when including only isospin-$1/2$ intermediate states reads 
\begin{equation}\label{wfr}
    Z_p = \frac{1}{1-\Sigma_p'}\Big|_{\slashed p = m}=1+\Sigma_p'+\mathcal{O}(p^3)|_{\slashed p = m}= 1+\frac{3 g_A^2 m_\pi^2}{32\pi^2F^2}\left(3 \log\left(\frac{m}{m_\pi}\right)-2\right),
\end{equation}
where $\Sigma_p$ is the self energy of the proton. Since we are considering the $\Delta(1232)$ as an intermediate state, we also have to take into account the additional self-energy loop that enters the wave-function renormalization. Also in this case, we took the $\mathcal{O}(p^2)$ term (there is a power-counting breaking term at $\mathcal{O}(p^0)$ which is subtracted) and added it to Eq.~\ref{wfr}.

We compare our model with the full set of data  of 
Refs.~\cite{Hornidge:2012ca} on the angular cross section
\begin{equation}
\frac{\mathrm{d}\sigma}{\mathrm{d}\Omega}=\frac{|\vec{q}|m^2}{2\pi W(s-m^2)}\sum_\epsilon{
\frac{\text{Tr}\left[
\mathcal{M}^*\cdot(\slashed{p}'+m)\cdot\mathcal{M}\cdot(\slashed{p}+m)
\right]}
{2}},
\end{equation}
where the sum over polarizabilities is taken because we are working with unpolarized photons, and with the linearly polarized photon asymmetry
\begin{equation}
\Sigma=\frac{\mathrm{d}\sigma_\perp-\mathrm{d}\sigma_\parallel}{\mathrm{d}\sigma_\perp+\mathrm{d}\sigma_\parallel},
\end{equation}
with $d\sigma_\perp$ and $d\sigma_\parallel$ the angular cross sections for photon polarization perpendicular and parallel to the reaction plane with the pion and the outgoing proton. 
In the CGLM representation, the differential cross section and photon asymmetry are written with the help of the response functions
\begin{align}
\nonumber R_T =& |\mathcal{F}_1|^2 + |\mathcal{F}_2|^2 + \frac12\sin^2\theta\left(|\mathcal{F}_3|^2+|\mathcal{F}_4|^2\right)
\\\nonumber&- \text{Re}\left[
2\cos\theta \mathcal{F}_1^*\mathcal{F}2- \sin^2\theta\left(
\mathcal{F}_1^*\mathcal{F}_4+\mathcal{F}_2^*\mathcal{F}_3+\cos\theta \mathcal{F}_3^*\mathcal{F}_4\right)\right],\\
\nonumber R_{TT} =& \frac12\sin^2\theta\left(|\mathcal{F}_3|^2+|\mathcal{F}_4|^2\right)
\\\nonumber&+ \text{Re}\left[ \sin^2\theta\left(
\mathcal{F}_1^*\mathcal{F}_4+\mathcal{F}_2^*\mathcal{F}_3+\cos\theta \mathcal{F}_3^*\mathcal{F}_4\right)\right],
\end{align}
with which one obtains
\begin{align}
\nonumber\frac{\mathrm{d}\sigma}{\mathrm{d}\Omega_\pi}=\frac{|\vec q|}{k_\gamma}R_T
\hspace{5mm}\text{and}\hspace{5mm}
\Sigma=-\frac{R_{TT}}{R_T}.
\end{align}
As for the lowest multipoles $E_{0+}$, $M_{1+}$, $M_{1-}$ and $E_{1+}$, they read~\cite{Bernard:1992nc}:
\begingroup
\renewcommand*{\arraystretch}{1.5}{
\begin{align*}
\left(
\begin{array}{c}
E_{0+}\\
M_{1+}\\
M_{1-}\\
E_{1+}
\end{array}\right)
=\bigint_{-1}^1\mathrm{d}x
\left(\begin{array}{cccc}
\frac{1}{2}P_0(x) & -\frac{1}{2}P_1(x) & 0 & \frac{1}{6}\left[P_0(x)-P_2(x)\right]\\
\frac{1}{4}P_1(x) & -\frac{1}{4}P_2(x) & \frac{1}{12}\left[P_2(x)-P_0(x)\right] & 0\\
-\frac{1}{2}P_1(x) & \frac{1}{2}P_0(x) & \frac{1}{6}\left[P_0(x)-P_2(x)\right] & 0\\
\frac{1}{4}P_1(x) & -\frac{1}{4}P_2(x) & \frac{1}{12}\left[P_0(x)-P_2(x)\right] & \frac{1}{10}\left[P_1(x)-P_3(x)\right]
\end{array}\right)
\left(
\begin{array}{c}
\mathcal{F}_1(x)\\
\mathcal{F}_2(x)\\
\mathcal{F}_3(x)\\
\mathcal{F}_4(x)
\end{array}\right),
\end{align*}
}\endgroup
where $x=\cos(\theta)$.

\section{Results and discussion}
Following the calculations performed in~\cite{Blin:2015}, we studied the neutral pion photoproduction on the proton in a fully covariant ChPT 
calculation and with the inclusion of the isospin-$3/2$ virtual states. There the obtained fits described the steep increase of the cross section 
with the photon energy very well, even at energies higher than 200~MeV. The low-energy constants obtained in the fits are shown in Table~\ref{tab:1}. 
One can see that the low-energy constant $g_A$ perfectly agrees with the expected result obtained from $\beta$-decay data. As for $\tilde{c}_{67}=c_6+c_7$, it 
corresponds to a combination of the nuclear magnetic moments, which is calculated in the $\overline{MS}$ scheme and with 
the inclusion of the $\Delta(1232)$ in~\cite{Ledwig:2011cx}. An analogous 
calculation in the $\widetilde{MS}$ scheme used in the present work yields an expected value of $\tilde{c}_{67}=2.5$, which is also very close to the free fit 
performed here. There are not yet any studies of $\tilde{d}_{89}=d_8+d_9$ in the EOMS scheme with $\Delta(1232)$ degrees of freedom. Therefore we 
let it completely free an obtain a very reasonable and natural value for its result. The low-energy constants $d_{16}$ and $d_{18}$ have been studied in the $\widetilde{MS}$ scheme with inclusion of $\Delta(1232)$ in~\cite{Alarcon:2012kn}. The couplings are the same as in the present work and the study is performed up to the same order. The constant $d_{16}$ is absorbed into the renormalization of $g_A$. Therefore our combined value $\tilde{d}_{168}=2d_{16}-d_{18}$ corresponds to a $d_{18}$ of around 7.6~MeV. This is interestingly quite different from the expected value in the above-mentioned work. Finally, another interesting fact is that, when letting $g_M$ be a free fitting constant, it assumes the value of $3.1$. A study performed in~\cite{Blin:2015era} showed that the expected value of $g_M$ from electromagnetic decay data of the $\Delta(1232)$ is $3.16$, which is in perfect agreement with the result from our fit.

\begin{table}
\begin{center}
\begin{tabular}{l||c|c|c|c|c}
  & $g_A$ & $\tilde{c}_{67}$ & $\tilde{d}_{89}$ [GeV$^{-2}$] & $\tilde{d}_{168} $ [GeV$^{-2}$]& $\chi^2$/d.o.f. \\ 
\hline\hline 
Full Model & {\bf 1.27} & 2.33 & 1.46 & -12.1 & 0.69 \\ 
\hline 
Full Model & 1.24 & 2.36 & 1.46 & -11.1 &  0.68
\end{tabular} 
\end{center}
\caption{LEC values in different versions of the model. Fixed values appear in boldface.}\label{tab:1}
\end{table}

As an extension of this work, we also test the convergence of the model, by proposing the calculation up to the 
next chiral order, which in the presented counting scheme corresponds to a calculation at $\mathcal{O}(p^{7/2})$. The additional contribution is expected to be 
small. With this model, it will then be possible to study observables like the differential cross sections, photon asymmetries and multipoles, among others. 
By fits to experimental data, one obtains further insight into the low-energy constants' values at the considered order, and therefore it is then 
possible to make more reliable statements about the consistency between these results and those from previous works.

\section*{Acknowledgements}
This research was supported by the Spanish Ministerio de Econom\'ia y Competitividad and European FEDER funds under Contracts No. FIS2011-28853-C02-01 and FIS2014-51948-C2-2-P, by Generalitat Valenciana under Contract No. PROMETEO/20090090 and by the EU HadronPhysics3 project, Grant Agreement No. 283286. A.N. Hiller Blin acknowledges support from the Santiago Grisol\'ia program of the  Generalitat Valenciana. We thank D. Hornidge
for providing us with the full set of data from Ref.~\cite{Hornidge:2012ca}.


\begin{thebibliography}{99}

%~\cite{DeBaenst:1971hp}
\bibitem{DeBaenst:1971hp}
  P.~De Baenst,
  %``An improvement on the kroll-ruderman theorem,''
  Nucl.\ Phys.\ B {\bf 24} (1970) 633.
  %%CITATION = NUPHA,B24,633;%%

%~\cite{Vainshtein:1972ih}
\bibitem{Vainshtein:1972ih}
  A.~I.~Vainshtein and V.~I.~Zakharov,
  %``Low-energy theorems for photoproduction and electropion production at threshold,''
  Nucl.\ Phys.\ B {\bf 36} (1972) 589.
  %%CITATION = NUPHA,B36,589;%%


%~\cite{Drechsel:1992pn}
\bibitem{Drechsel:1992pn} 
  D.~Drechsel and L.~Tiator,
  %``Threshold pion photoproduction on nucleons,''
  J.\ Phys.\ G {\bf 18}, 449 (1992).
  %%CITATION = JPHGB,G18,449;%%
  %162 citations counted in INSPIRE as of 30 Jun 2014

%~\cite{Bernard:2006gx}
\bibitem{Bernard:2006gx} 
  V.~Bernard and U.~-G.~Meissner,
  %``Chiral perturbation theory,''
  Ann.\ Rev.\ Nucl.\ Part.\ Sci.\  {\bf 57}, 33 (2007).
  %%CITATION = HEP-PH/0611231;%%
  %87 citations counted in INSPIRE as of 30 Jun 2014  

%~\cite{Gasparyan:2010xz}
\bibitem{Gasparyan:2010xz} 
  A.~Gasparyan and M.~F.~M.~Lutz,
  Nucl.\ Phys.\ A {\bf 848}, 126 (2010).  


%~\cite{Mazzucato:1986dz}
\bibitem{Mazzucato:1986dz} 
  E.~Mazzucato, P.~Argan, G.~Audit, A.~Bloch, N.~de Botton, N.~d'Hose, J.~L.~Faure and M.~L.~Ghedira {\it et al.},
  %``A Precise Measurement of Neutral Pion Photoproduction on the Proton Near Threshold,''
  Phys.\ Rev.\ Lett.\  {\bf 57}, 3144 (1986).
  %%CITATION = PRLTA,57,3144;%%


%~\cite{Beck:1990da}
\bibitem{Beck:1990da}
  R.~Beck, F.~Kalleicher, B.~Schoch, J.~Vogt, G.~Koch, H.~Stroher, V.~Metag and J.~C.~McGeorge {\it et al.},
  %``Measurement of the p (gamma, pi0) cross-section at threshold,''
  Phys.\ Rev.\ Lett.\  {\bf 65} (1990) 1841.
  %%CITATION = PRLTA,65,1841;%%



%~\cite{Bernard:1991rt}
\bibitem{Bernard:1991rt}
  V.~Bernard, N.~Kaiser, J.~Gasser and U.~G.~Meissner,
  %``Neutral pion photoproduction at threshold,''
  Phys.\ Lett.\ B {\bf 268} (1991) 291.
  %%CITATION = PHLTA,B268,291;%%

%~\cite{Bernard:1992nc}
\bibitem{Bernard:1992nc}
  V.~Bernard, N.~Kaiser and U.~G.~Meissner,
  %``Threshold pion photoproduction in chiral perturbation theory,''
  Nucl.\ Phys.\ B {\bf 383} (1992) 442.
  %%CITATION = NUPHA,B383,442;%%


%~\cite{Bernard:1994gm}
\bibitem{Bernard:1994gm} 
  V.~Bernard, N.~Kaiser and U.~-G.~Meissner,
  %``Neutral pion photoproduction off nucleons revisited,''
  Z.\ Phys.\ C {\bf 70}, 483 (1996).
  %%CITATION = HEP-PH/9411287;%%
  %112 citations counted in INSPIRE as of 23 May 2014

%~\cite{Bernard:1995cj}
\bibitem{Bernard:1995cj}
  V.~Bernard, N.~Kaiser and U.~G.~Meissner,
  %``Chiral symmetry and the reaction gamma p ---> pi0 p,''
  Phys.\ Lett.\ B {\bf 378} (1996) 337.
  %%CITATION = HEP-PH/9512234;%%
  %63 citations counted in INSPIRE as of 31 Oct 2014
%
%
%~\cite{Bernard:2001gz}
\bibitem{Bernard:2001gz} 
  V.~Bernard, N.~Kaiser and U.~-G.~Meissner,
  %``Aspects of near threshold neutral pion photoproduction off protons,''
  Eur.\ Phys.\ J.\ A {\bf 11}, 209 (2001).
  %%CITATION = HEP-PH/0102066;%%
  %48 citations counted in INSPIRE as of 23 May 2014

%~\cite{Hornidge:2012ca}
\bibitem{Hornidge:2012ca} 
  D.~Hornidge {\it et al.}  [A2 and CB-TAPS Collaborations],
  %``Accurate Test of Chiral Dynamics in the γ→p→π0p Reaction,''
  Phys.\ Rev.\ Lett.\  {\bf 111}, no. 6, 062004 (2013).
  %%CITATION = ARXIV:1211.5495;%%
  %11 citations counted in INSPIRE as of 23 May 2014


%
%~\cite{FernandezRamirez:2012nw}
\bibitem{FernandezRamirez:2012nw} 
  C.~Fernandez-Ramirez and A.~M.~Bernstein,
  %``Upper Energy Limit of Heavy Baryon Chiral Perturbation Theory in Neutral Pion Photoproduction,''
  Phys.\ Lett.\ B {\bf 724}, 253 (2013).
  %%CITATION = ARXIV:1212.3237;%%
  %11 citations counted in INSPIRE as of 23 May 2014

%
%~\cite{Hilt:2013uf}
\bibitem{Hilt:2013uf} 
  M.~Hilt, S.~Scherer and L.~Tiator,
  %``Threshold $\pi^0$ photoproduction in relativistic chiral perturbation theory,''
  Phys.\ Rev.\ C {\bf 87}, no. 4, 045204 (2013).
  %%CITATION = ARXIV:1301.5576;%%
  %11 citations counted in INSPIRE as of 23 May 2014


%~\cite{Ericson:1988gk}
\bibitem{Ericson:1988gk}
  T.~E.~O.~Ericson and W.~Weise,
  %``Pions And Nuclei,''
  OXFORD, UK: CLARENDON (1988) 479 P. (THE INTERNATIONAL SERIES OF MONOGRAPHS ON PHYSICS, 74)



%~\cite{Hemmert:1996xg}
\bibitem{Hemmert:1996xg} 
  T.~R.~Hemmert, B.~R.~Holstein and J.~Kambor,
  %``Systematic 1/M expansion for spin 3/2 particles in baryon chiral perturbation theory,''
  Phys.\ Lett.\ B {\bf 395}, 89 (1997).
  %%CITATION = HEP-PH/9606456;%%
  %115 citations counted in INSPIRE as of 02 Jul 2014




%~\cite{Pascalutsa:2006up}
\bibitem{Pascalutsa:2006up}
  V.~Pascalutsa, M.~Vanderhaeghen and S.~N.~Yang,
  %``Electromagnetic excitation of the Delta(1232)-resonance,''
  Phys.\ Rept.\  {\bf 437} (2007) 125.
  %%CITATION = HEP-PH/0609004;%%
  %167 citations counted in INSPIRE as of 02 Jul 2014

%~\cite{Chew:1957tf}
\bibitem{Chew:1957tf}
  G.~F.~Chew, M.~L.~Goldberger, F.~E.~Low and Y.~Nambu,
  %``Relativistic dispersion relation approach to photomeson production,''
  Phys.\ Rev.\  {\bf 106} (1957) 1345.
  %%CITATION = PHRVA,106,1345;%%
  %540 citations counted in INSPIRE as of 18 Nov 2014

%~\cite{Adler:1968tw}
\bibitem{Adler:1968tw}
  S.~L.~Adler,
  %``Photoproduction, electroproduction and weak single pion production in the (3,3) resonance region,''
  Annals Phys.\  {\bf 50} (1968) 189.
  %%CITATION = APNYA,50,189;%%
  %351 citations counted in INSPIRE as of 18 Nov 2014


%~\cite{Pascalutsa:2004pk}
\bibitem{Pascalutsa:2004pk} 
  V.~Pascalutsa and J.~A.~Tjon,
  %``Pion photoproduction on nucleons in a covariant hadron-exchange model,''
  Phys.\ Rev.\ C {\bf 70}, 035209 (2004).
  %%CITATION = NUCL-TH/0407068;%%
  %24 citations counted in INSPIRE as of 02 Jul 2014

%~\cite{Pascalutsa:2005vq}
\bibitem{Pascalutsa:2005vq} 
  V.~Pascalutsa and M.~Vanderhaeghen,
  %``Chiral effective-field theory in the Delta(1232) region: I. Pion electroproduction on the nucleon,''
  Phys.\ Rev.\ D {\bf 73}, 034003 (2006).
  %%CITATION = HEP-PH/0512244;%%
  %65 citations counted in INSPIRE as of 02 Jul 2014
% interesting, resonance region
% for some reason they don't include all diagrams.



%~\cite{FernandezRamirez:2005iv}
\bibitem{FernandezRamirez:2005iv}
  C.~Fernandez-Ramirez, E.~Moya de Guerra and J.~M.~Udias,
  %``Effective Lagrangian approach to pion photoproduction from the nucleon,''
  Annals Phys.\  {\bf 321} (2006) 1408.
  %%CITATION = NUCL-TH/0509020;%%

%~\cite{Alarcon:2012kn}
\bibitem{Alarcon:2012kn}
  J.~M.~Alarcon, J.~Martin Camalich and J.~A.~Oller,
  %``Improved description of the $\pi N$-scattering phenomenology in covariant baryon chiral perturbation theory,''
  Annals Phys.\  {\bf 336} (2013) 413.
  %%CITATION = ARXIV:1210.4450;%%


%~\cite{Blin:2015} 
\bibitem{Blin:2015} 
 A.~N.~Hiller Blin, T.~Ledwig and M.~J.~Vicente Vacas,
 %``Chiral dynamics in the $\overrightarrow{γ}p → pπ^0$ reaction,"
 Phys.\ Lett.\ B {\bf 747} (2015) 217.

%~\cite{Lensky:2009uv}
\bibitem{Lensky:2009uv}
  V.~Lensky and V.~Pascalutsa,
  %``Predictive powers of chiral perturbation theory in Compton scattering off protons,''
  Eur.\ Phys.\ J.\ C {\bf 65} (2010) 195.
  %%CITATION = ARXIV:0907.0451;%%
  %51 citations counted in INSPIRE as of 13 Nov 2014
%
%


%~\cite{Fettes:2000gb}
\bibitem{Fettes:2000gb} 
  N.~Fettes, U.~G.~Meissner, M.~Mojzis and S.~Steininger,
  %``The Chiral effective pion nucleon Lagrangian of order p**4,''
  Annals Phys.\  {\bf 283}, 273 (2000)
  [Erratum-ibid.\  {\bf 288}, 249 (2001)]
  %%CITATION = HEP-PH/0001308;%%


%~\cite{Pascalutsa:2002pi}
\bibitem{Pascalutsa:2002pi} 
  V.~Pascalutsa and D.~R.~Phillips,
  %``Effective theory of the delta(1232) in Compton scattering off the nucleon,''
  Phys.\ Rev.\ C {\bf 67}, 055202 (2003).
  %%CITATION = NUCL-TH/0212024;%%
  %119 citations counted in INSPIRE as of 02 Jul 2014


%~\cite{Gegelia:1999gf}
\bibitem{Gegelia:1999gf}
  J.~Gegelia and G.~Japaridze,
  %``Matching heavy particle approach to relativistic theory,''
  Phys.\ Rev.\ D {\bf 60} (1999) 114038.
  %%CITATION = HEP-PH/9908377;%%
  %96 citations counted in INSPIRE as of 31 Oct 2014

%~\cite{Fuchs:2003qc}
\bibitem{Fuchs:2003qc}
  T.~Fuchs, J.~Gegelia, G.~Japaridze and S.~Scherer,
  %``Renormalization of relativistic baryon chiral perturbation theory and power counting,''
  Phys.\ Rev.\ D {\bf 68} (2003) 056005.
  %%CITATION = HEP-PH/0302117;%%
  %167 citations counted in INSPIRE as of 31 Oct 2014

%~\cite{Fuchs:2003ir}
\bibitem{Fuchs:2003ir}
  T.~Fuchs, J.~Gegelia and S.~Scherer,
  %``Electromagnetic form-factors of the nucleon in relativistic baryon chiral perturbation theory,''
  J.\ Phys.\ G {\bf 30} (2004) 1407.
  %%CITATION = NUCL-TH/0305070;%%

%~\cite{Lehnhart:2004vi}
\bibitem{Lehnhart:2004vi}
  B.~C.~Lehnhart, J.~Gegelia and S.~Scherer,
  %``Baryon masses and nucleon sigma terms in manifestly Lorentz-invariant baryon chiral perturbation theory,''
  J.\ Phys.\ G {\bf 31} (2005) 89.
  %%CITATION = HEP-PH/0412092;%%

%~\cite{Schindler:2006it}
\bibitem{Schindler:2006it}
  M.~R.~Schindler, T.~Fuchs, J.~Gegelia and S.~Scherer,
  %``Axial, induced pseudoscalar, and pion-nucleon form-factors in manifestly Lorentz-invariant chiral perturbation theory,''
  Phys.\ Rev.\ C {\bf 75} (2007) 025202.
  %%CITATION = NUCL-TH/0611083;%%

%~\cite{Schindler:2006ha}
\bibitem{Schindler:2006ha}
  M.~R.~Schindler, D.~Djukanovic, J.~Gegelia and S.~Scherer,
  %``Chiral expansion of the nucleon mass to order(q**6),''
  Phys.\ Lett.\ B {\bf 649} (2007) 390.
  %%CITATION = HEP-PH/0612164;%%


%~\cite{Geng:2008mf}
\bibitem{Geng:2008mf}
  L.~S.~Geng, J.~Martin Camalich, L.~Alvarez-Ruso and M.~J.~Vicente Vacas,
  %``Leading SU(3)-breaking corrections to the baryon magnetic moments in Chiral Perturbation Theory,''
  Phys.\ Rev.\ Lett.\  {\bf 101} (2008) 222002.
  %%CITATION = ARXIV:0805.1419;%%
  %57 citations counted in INSPIRE as of 31 Oct 2014

%~\cite{Geng:2009ik}
\bibitem{Geng:2009ik}
  L.~S.~Geng, J.~Martin Camalich and M.~J.~Vicente Vacas,
  %``SU(3)-breaking corrections to the hyperon vector coupling f(1)(0) in covariant baryon chiral perturbation theory,''
  Phys.\ Rev.\ D {\bf 79} (2009) 094022.
  %%CITATION = ARXIV:0903.4869;%%
  %24 citations counted in INSPIRE as of 31 Oct 2014

%~\cite{MartinCamalich:2010fp}
\bibitem{MartinCamalich:2010fp}
  J.~Martin Camalich, L.~S.~Geng and M.~J.~Vicente Vacas,
  %``The lowest-lying baryon masses in covariant SU(3)-flavor chiral perturbation theory,''
  Phys.\ Rev.\ D {\bf 82} (2010) 074504.
  %%CITATION = ARXIV:1003.1929;%%
  %54 citations counted in INSPIRE as of 31 Oct 2014

%~\cite{Alarcon:2011zs}
\bibitem{Alarcon:2011zs}
  J.~M.~Alarcon, J.~Martin Camalich and J.~A.~Oller,
  %``The chiral representation of the $\pi N$ scattering amplitude and the pion-nucleon sigma term,''
  Phys.\ Rev.\ D {\bf 85} (2012) 051503.
  %%CITATION = ARXIV:1110.3797;%%
  %80 citations counted in INSPIRE as of 31 Oct 2014

%~\cite{Ledwig:2011cx}
\bibitem{Ledwig:2011cx}
  T.~Ledwig, J.~M.~Camalich, V.~Pascalutsa and M.~Vanderhaeghen,
  %``The Nucleon and $\Delta$(1232) form factors at low momentum-transfer and small pion masses,''
  Phys.\ Rev.\ D {\bf 85} (2012) 034013.
  %%CITATION = ARXIV:1405.5456;%%

%~\cite{Chen:2012nx}
\bibitem{Chen:2012nx}
  Y.~H.~Chen, D.~L.~Yao and H.~Q.~Zheng,
  %``Analyses of pion-nucleon elastic scattering amplitudes up to $\mathcal{O}(p^4)$ in extended-on-mass-shell subtraction scheme,''
  Phys.\ Rev.\ D {\bf 87} (2013) 5,  054019.
  %%CITATION = ARXIV:1212.1893;%%
  %10 citations counted in INSPIRE as of 31 Oct 2014

%~\cite{Scherer:2012xha}
\bibitem{Scherer:2012xha}
  S.~Scherer and M.~R.~Schindler,
  %``A Primer for Chiral Perturbation Theory,''
  Lect.\ Notes Phys.\  {\bf 830} (2012) pp.1.
  %%CITATION = LNPHA,830,pp.1;%%


%~\cite{Alvarez-Ruso:2013fza}
\bibitem{Alvarez-Ruso:2013fza}
  L.~Alvarez-Ruso, T.~Ledwig, J.~Martin Camalich and M.~J.~Vicente Vacas,
  %``Nucleon mass and pion-nucleon sigma term from a chiral analysis of lattice QCD data,''
  Phys.\ Rev.\ D {\bf 88} (2013) 5,  054507.

%~\cite{Ledwig:2014rfa}
\bibitem{Ledwig:2014rfa}
  T.~Ledwig, J.~M.~Camalich, L.~S.~Geng and M.~J.~V.~Vacas,
  %``Octet-baryon axial-vector charges and SU(3)-breaking effects in the semileptonic hyperon decays,''
  Phys.\ Rev.\ D {\bf 90} (2014) 054502.
  %%CITATION = ARXIV:1405.5456;%%

%\cite{Blin:2015era}
\bibitem{Blin:2015era}
  A.~Hiller Blin, Th.~Gutsche, T.~Ledwig and V.~E.~Lyubovitskij,
  %``Hyperon forward spin polarizability gamma0 in baryon chiral perturbation theory,''
  [arXiv:hep-ph/1509.00955].
  %%CITATION = ARXIV:1401.0140;%%
\end{thebibliography}
\end{document}